\documentclass[%
superscriptaddress,
nobibnotes,
amsmath,amssymb,
aps,
floatfix,
]{revtex4-2}

\usepackage{subfig}
\usepackage{graphicx}
\usepackage{dcolumn}
\usepackage{bm}
\usepackage{color} 	
\usepackage{textcomp}
\usepackage{tikz}
\usepackage[colorlinks=true,linkcolor=blue]{hyperref}%

 \newcommand{\dx}{\frac{{\rm d} }{{\rm d} x}}
 
\providecommand\bcdot{\boldsymbol{\cdot}}

\providecommand\bcdot{\pmb{\cdot}}

\newcommand\mbfq{\mathbf{q}}
\newcommand\mbfpi{\mathbf{\Pi}}
\newcommand\mbfi{\mathbf{I}}
\newcommand\mbfx{\mathbf{x}}
\newcommand\bR{{R}}
\newcommand{\la}{\lambda}
\newcommand{\ka}{\kappa}
\newcommand{\kam}{\kappa_m}

\newcommand{\redline}{\raisebox{2pt}{\tikz{\draw[-,red,solid,line width = 1.5pt](0,0) -- (1.75,0);}}}
\newcommand{\dashed}{\tikz[baseline=-0.6ex]\draw[thick,dashed,line width = 1.5pt] (0,0)--(1.75,0);}
\newcommand{\blackline}{\raisebox{2pt}{\tikz{\draw[-,black,solid,line width = 1.5pt](0,0) -- (1.75,0);}}}
\newcommand{\longdashed}{\tikz[baseline=-0.6ex]\draw [thick,dash pattern={on 10pt off 4pt},line width = 1.5pt] (0,0) -- (1.75,0);}
\newcommand{\dashdot}{\tikz[baseline=-0.6ex]\draw [thick,dash pattern={on 10pt off 3pt on 1pt off 3pt},line width = 1.5pt] (0,0) -- (1.75,0);}
\newcommand{\dashdotdot}{\tikz[baseline=-0.6ex]\draw[thick,dash pattern={on 10pt off 3pt on 1pt off 3pt on 1pt off 3pt},line width = 1.5pt] (0,0)--(1.75,0);}
\newcommand{\bluecircle}{\tikz{\filldraw[blue] (0,0) circle (3.0pt)}}
\newcommand{\opencircle}{\tikz{\draw[black] (0,0) circle (3.0pt)}}
\newcommand{\filledsquare}{\tikz{\filldraw[darkgreen] (0,0) rectangle (0.25,0.25)}}
\newcommand{\filledtriangle}{\tikz{\filldraw[blue] (-0.15,0) -- (0.15,0) -- (0,0.2)}}

\newcommand{\filleddaimond}{\tikz{\filldraw[black] (0.15,0) -- (0,0.15) -- (-0.15,0) -- (0,-0.15)}}
\definecolor{orange}{rgb}{1,0.5,0}
\definecolor{pinegreen}{rgb}{0,1,0.95}
\definecolor{darkgreen}{rgb}{0,0.54,0}
\definecolor{gray}{rgb}{0.3, 0.3, 0.3}
\begin{document}

\preprint{APS/123-QED}

\title{Effects of Molecular Diffusivity on Shock Wave Structures in Monatomic Gases}

\author{M.~H.~Lakshminarayana Reddy}
\email{l.mh@hw.ac.uk}
\affiliation{School of Engineering and Physical Sciences, Heriot-Watt University, Edinburgh EH14 4AS, Scotland, UK.}

\author{S.~Kokou~Dadzie}
\email{k.dadzie@hw.ac.uk}
\affiliation{School of Engineering and Physical Sciences, Heriot-Watt University, Edinburgh EH14 4AS, Scotland, UK.}

\date{\today}

\begin{abstract}
We present a full investigation into shock wave profile description using hydrodynamics models.
We identified constitutive equations that provide better agreement for all parameters involved in testing hydrodynamic equations for the prediction of shock structure in a monatomic gas in the Mach number range $1.0-11.0$. The constitutive equations are extracted from a previously derived thermomechanically  consistent Burnett regime continuum flow model. The numerical computations of the resulting hydrodynamic equations along with classical ones are performed using a finite difference global solution (FDGS) scheme. Compared to previous studies that focussed mainly on the density profile across the shock, here we also include temperature profiles as well as non-negativity of entropy production throughout the shock. The results obtained show an improvement upon those obtained previously in the bi-velocity (or volume/mass diffusion) hydrodynamics and are more accurate than in the hydrodynamic models from expansions method solutions to the Boltzmann equation.
\end{abstract}

\pacs{Valid PACS appear here}

\maketitle



\section{\label{sec:I}Introduction}

Shock wave structure description is one of the best-known example and a simple highly non-equilibrium compressible flow problem (where large gradients of hydrodynamic fields are present). It has been subjected to theoretical, numerical and experimental attacks mainly from around the middle of the twentieth century \cite{CF1948,LR1957,Grad1952,Bird1994,Reeseetal1995,LeVeque2002,Greenshields2007,RA2015,Reddy2016,RA2016} due to the advantages it provides: (i) flow is one-dimensional and steady state; (ii) no solid boundaries; and (iii) the upstream and downstream states are in equilibrium and are connected by simple laws/relations (the Rankine-Hugoniot relations).
Theoretical and numerical studies of the shock structure based on the classical Navier-Stokes (NS) equations are described in the literature \cite{CF1948,Mises1950,GP1953,LR1957,NarasimhaDas1990,PEM1991}. In addition, accurate shock density measurements have been carried out and reported for argon and nitrogen gases with Mach number that ranges from supersonic to hypersonic by \citet{Alsmeyer1976}. A number of experimental studies \cite{LH1963,Camac1965,Sf1965,Russell1965,RT1966,Schmidt1969,Rieutord1970,Garenetal1974} were reported prior to  \citet{Alsmeyer1976} but most of them are reassembled in Alsmeyer's work for comparison. Eventually, it has been recognized that  shock structures in monatomic gases are not well described by the Navier-Stokes theory. The shock thickness predicted is too small compared to experiments for Mach numbers larger than  approximately $1.5$. In fact, the inadequacy of the classical Navier-Stokes equations in describing some compressible flows has been the subject of discussion from a long time \cite{Alsmeyer1976,Sone2000,Greenshields2007,Durst2008}. The failure of the equations in a shock structure description may be tied up to the basic assumptions such as linear constitutive relations represented by Newton's law of viscosity and Fourier's law of heat conduction used in closing the system \cite{Stokes1845}  and / or breakdown of continuum assumption as the mean free path becomes comparable to the characteristic length scale of the system. In other words one can say that the failure can be attributed to basic inherent limitations such as the breakdown of local equilibrium hypothesis.

The principal parameter which is often used to classify the non-equilibrium state of a gas flow is the Knudsen number, $\rm{Kn}$, and is defined as the ratio of the mean free path of the gas molecules to the characteristic length of the flow system. $\rm{Kn}$ characterizes the gas rarefaction which means that it measures departure from the local equilibrium. Continuum assumption is valid for vanishing Knudsen numbers where the gas can be  assumed to undergo a large number of collisions over the typical length scale. As $\rm{Kn}$ increases the notion of the gas as a continuum fluid becomes less valid and the departure of the gas from the local thermodynamic equilibrium increases. Therefore, the range of use of the continuum-equilibrium assumption is limited and confined to $\rm{Kn} \lesssim 0.01$. Generally, the shock macroscopic parameter called the shock thickness is related to the Knudsen number and typically falls between $\approx 0.2$ and $\approx 0.3$ \cite{Greenshields2007}. Clearly, the range of $\rm{Kn}$ found in the shock problem is beyond the classical continuum-$\rm{Kn}$ regime and falls into the so-called `intermediate-$\rm{Kn}$' regime ($0.01 \lesssim \rm{Kn} \lesssim 1$). Deriving appropriate continuum hydrodynamic models or improving the range of applicability of the existing ones (the Navier-Stokes equations) beyond their limits into the so-called `intermediate-$\rm{Kn}$' regime ($0.01 \lesssim \rm{Kn} \lesssim 1$) is still is a critical active area of research.

Gas flows may be described at any Knudsen number ($\rm{Kn}$)  by the Boltzmann equation (the central equation in kinetic theory of gases) \cite{Boltzmann1878}. The solution to the Boltzmann equation via particle based method DSMC (direct simulation Monte Carlo) technique has been found to be very useful in rarefied regime \cite{Bird1970a,Bird1970b,Bird1994}. Meanwhile, there are significant attempts made to formulate solution techniques based on the extended hydrodynamic approach. In an “extended” hydrodynamic approach the problem of directly solving the Boltzmann equation is replaced by solving a system of generalized transport equations either in terms of the extended hydrodynamic variable set or in terms of higher order space derivatives of the hydrodynamic quantities \cite{Grad1949,CC1970,Cercignani1975,GK1994,HS2005}. These equations are typically obtained from the Boltzmann equation  by performing or using techniques such as: (i) the Chapman-Enskog (CE) expansion~\cite{Enskog1917,CC1970,Cercignani1975,HS2005}; (ii) Grad's moment method \cite{Grad1949,Grad1952,RG2019}; (iii) Maximum entropy method (also called Maximum probability principle sometimes)~\cite{Kogan1969}; and  iv) an hybrid method which combines aspect of Grad and CE method~\citep{GK1994,LM1996}. Following these solution techniques in extended hydrodynamic approach, numerous higher-order extended hydrodynamic models arose and most highlighted commonly used and referenced include: Grad's moment equations \cite{Grad1949,Grad1952,RG2019,Reddyetal2014} and Burnett equations \cite{CC1970}. All these higher-order extended hydrodynamic equations may be superior to some extent in describing the flow physics better than the classical NS equations; however, these equations come with some serious drawbacks with leading reason being that they are all invaded by physical and numerical instabilities and produce non-physical flow solutions in many cases \cite{Weiss1995,TS2004,Reddy2016}. To overcome these disadvantages in the extended hydrodynamic models, variants of Burnett equations (reduced/augmented Burnett \cite{LC1992}, BGK-Burnett \cite{Balakrishnan2004}, regularized Burnett\cite{JS2001}), regularized moment equations \cite{TS2004,HS2005,RA2020,RA2016a} and second-order descriptions and theories of \citet{Woods1993,Reeseetal1995} are proposed in the literature.

Recent notable works on improving shock structure prediction results over the classical Navier-Stokes may include: a second-order continuum theory of \citet{Paolucci2018}, a linear irreversible thermodynamic model of \citet{VelascoUribe2019}, recast Navier-Stokes of \citet{RD2020}, Onsager-Burnett equations of \citet{JadhavAgrawal2020}. These previous works paid less attention to temperature profile description across the shock layer. Recently, the authors reinterpreted shock structure predictions of the classical Navier-Stokes equations using a change of velocity variable \cite{RD2020}. The results on the shock density profiles and shock thicknesses better agreed with the experimental data. However, the procedure predicted very less values for density asymmetry factor and not so accurate prediction of the density-temperature separation distance. The present work expands on this previous work to identify constitutive relations with a full assessment of the shock structure problem that includes comparison of temperature profiles, density-temperature separation distance and also considering non-negativity of entropy production across the shock.

The paper is organized as follows. In Sec.~\ref{sec:II} we start with a brief overview of the classical Navier-Stokes equations for compressible flows and the modified constitutive relations are presented. In Sec.~\ref{sec:III} the modified Navier-Stokes equations are considered subject to shock structure problem in monatomic argon gas. The detail of the formulation of the problem and numerical procedure are then given. Sec.~\ref{sec:IV} is completely devoted to analysis based on comparison of shock macroscopic profiles and different shock macroscopic parameters with available experiments and other simulation data. Sec.~\ref{sec:V} is committed to evaluating the non-negativity of entropy generation within the shock layer. Finally, conclusions are drawn in Sec.~\ref{sec:VI}.

\section{\label{sec:II}{The modified continuum flow equations}}
We adopt the classical conservation equations in an Eulerian reference frame as given by:

mass balance equation
\begin{equation}
\label{eqn_mass}
\frac{\partial \rho}{\partial t} + \nabla \bcdot [ \rho\, U ]  = 0,
\end{equation}

momentum balance equation
\begin{equation}
\label{eqn_momentum}
\frac{\partial  \rho \, U }{\partial t}  +  \, \nabla \bcdot [ \rho \,U\otimes\,U\,+ \,p \,\mbfi \,+ \,\mbfpi ] = 0,
\end{equation}

energy balance equation
\begin{eqnarray}
\label{eqn_energy}
&& \frac{\partial}{\partial t} [\frac{1}{2} \rho \,U^2 + \rho \,e_{in}] + \nabla \bcdot [\frac{1}{2} \rho\, U^2\, U + \rho\, e_{in}\, U\,+\,(p \,\mbfi + \mbfpi) \bcdot U\,+\, \mbfq ] = 0,
\end{eqnarray}
where $\rho$ is the mass-density of the fluid, $U$ is the flow mass velocity, $p$ is the hydrostatic pressure, $e_{in}$ is the specific internal energy of the fluid, $\mbfpi$ is the shear stress tensor, $\mbfi$ is the identity tensor and $\mbfq$ is the heat flux vector. All these hydrodynamic fields are functions of time $t$ and spatial variable $\mbfx$. Additionally, $\nabla$ and $\nabla \bcdot$ denote the usual spatial gradient and divergence operators, respectively, while the operator $\otimes$ denotes the usual tensor product of two vectors. Expression for the specific internal energy is given by, $e_{in} = p / \rho (\gamma -1)$ with $\gamma$ being the isentropic exponent.
The constitutive models for the shear stress $\mbfpi$ and the heat flux vector $\mbfq$ as due to the Newton's law and the Fourier's law, are given respectively by,
\begin{align}
\mbfpi^{(NS)} &= -2 \,\mu \left[ \frac{1}{2} (\nabla U \,+ \,\nabla U')\, -\, \frac{1}{3}\, \mbfi \,\left(\nabla \bcdot U \right) \right] = -2\, \mu\, \mathring{\overline{\nabla U}}, \label{eqn_NSstress}\\
\mbfq^{(NS)} &= - \ka \, \nabla T, \label{eqn_NShf}
\end{align}
where $\nabla U'$ represents the transpose of $\nabla U$. Coefficients $\mu$ and $\kappa$ are the dynamic viscosity and  the heat conductivity, respectively. The shear stress can be expressed in terms of the symmetric part of the velocity gradient $\mathbf{D}\left(U\right) = (\nabla U \,+ \,\nabla U')/2$ and the divergence of the velocity field as
\begin{align}
\mbfpi^{(NS)} &= -2 \mu \left[ \mathbf{D}\left(U\right)\,-\,\frac{1}{3} \left( \nabla \bcdot U\right)\, \mbfi \right]
= -\,2 \,\mu \,\mathbf{D}\left(U\right)\,-\,\lambda\, \left( \nabla \bcdot U\right)\, \mbfi,
\end{align}
where $\lambda = -\frac{2}{3}\mu$ is the bulk-viscosity coefficient.

The system \eqref{eqn_mass} - \eqref{eqn_NShf} is the well known and widely accepted conventional fluid flow hydrodynamic model for a viscous and heat conducting fluid, called the classical Navier-Stokes equations. In the limit of vanishing viscous and heat conducting terms, the model reduced to the simple gas dynamics model known as Euler equations, which are used to model inviscid and non-diffusive flows. 


In the present study, to investigate our shock structure problem we adapt constitutive equations from previous studies \citet{Dadzie2013,Brenner2012}:
\begin{align}
\mbfpi &= -2\, \mu\, \mathring{\overline{\nabla U}} - 2\,\mu \, \mathring{\overline{\nabla J_D}} , \label{eqn_MNSstress}\\
\mbfq &= - \ka \, \nabla T - \frac{\gamma}{(\gamma - 1) {\rm Pr}}\,p\, J_D , \label{eqn_MNShf}
\end{align}
with
\begin{align}
J_D &= {\rm \kappa_m} \,\nabla \ln \rho, \label{eqn_diffusiveterms}
\end{align}
where $\rm{Pr}$ is the Prandtl number and ${\rm \kappa_m}$ is an additional transport coefficient, the molecular diffusivity coefficient and is related to the kinematic viscosity coefficient through the following relation:
\begin{equation}
\label{eqn_kappam}
{\rm \kappa_m} = {\rm \kappa_{m_0}} \frac{\mu}{\rho}, 
\end{equation}
where $\kappa_{m_0}$ is a positive constant. Using \eqref{eqn_diffusiveterms}, the fully modified stress in equation (\ref{eqn_MNSstress}) can be expressed as:
\begin{align}
\mbfpi &= -\,2 \,\mu \,\mathbf{D}\left(U\right)\,-\,\lambda\, \left( \nabla \bcdot U\right)\, \mbfi\, -\,2 \,\mu \, {\rm \kappa_m} \,\mathbf{D}\left(\nabla \ln \rho \right)\,-\,\lambda\, {\rm \kappa_m}\, \left( \nabla \bcdot \nabla \ln \rho \right)\, \mbfi.
\end{align}
Constitutive relations \eqref{eqn_MNSstress} - \eqref{eqn_diffusiveterms} are formally those proposed in \citet{Dadzie2013,Brenner2012} and used in \citet{Greenshields2007}. The difference in the current expression being the identified additional factor of $\gamma/((\gamma - 1) {\rm Pr})$ in the heat flux relation. In these constitutive equations, the additional components in both the shear stress and heat flux stem from the inclusion of molecular level diffusion in the construction of the full continuum flow model. They may therefore be referred to as volume/mass diffusion corrections to momentum and heat transport and were previously shown to improve the prediction of some non-equilibrium effects \cite{DadzieChariton2016,CharitonDadzie2017}. Examination of the linear stability of the continuum flow model closed with constitutive relations \eqref{eqn_MNSstress} - \eqref{eqn_diffusiveterms} to small perturbations following the procedure described in \citet{DadzieReese2010, Dadzie2013, RG2019, Reddyetal2019} revealed that they may become temporally or spatially unstable for some values of ${\rm k_{m_0}}$. However, these  constitutive equations are shown to be temporally and spatially stable for any value of ${\rm k_{m_0}}$ in a fully thermomechanically consistent set of equations where an additional transport equation is added in their derivation \citet{Dadzie2013}.   

In the next sections, we show that use of constitutive relations \eqref{eqn_MNSstress} - \eqref{eqn_diffusiveterms} considerably improve predictions in the shock profile problem compared to previous models.

\section{\label{sec:III} The shock structure problem in a monatomic gas}

A shock wave is generated when a supersonic gas flows into a subsonic gas; mathematically, this is nothing but a discontinuity across which the hydrodynamic fields undergo discontinuous jumps. In other words, a shock wave involves a transition between a uniform upstream flow and a uniform downstream flow, thus, we can treat the shock wave as an interface of finite thickness between two different equilibrium states of a gas. Due to the interaction with the subsonic gas particles, the supersonic gas particles are slowed down and causes a sharp increase in the density and temperature at this point. For instance, the normal shock wave can be easily visualized in a balloon bursting \cite{Mackenzie2006,Reddy2016}: when a balloon bursts, the interior gas is expelled outward radially and it collides with the stationary exterior gas and causes a build up of particles at the boundary between the two gases, which moves radially outward.

The evolution of a monatomic ideal gas flow is determined by the density $\rho$, the velocity $U$ and the temperature $T$ at any point in space and time. Its pressure $p$ obeys the perfect gas law,
\begin{equation}
\label{eqn_S1}
 p = \rho \, \bR \, T,
\end{equation}
Where $\bR = {\rm k_B}/m$ is the specific gas constant with ${\rm k_B}$ and $m$ being the Boltzmann constant and  the molecular mass, respectively. In terms of the specific heat at constant pressure, $c_p$, and constant volume, $c_v$, a monatomic ideal gas is characterized by
\begin{equation}
\label{eqn_S2}
 c_p = \frac{\gamma}{(\gamma-1)} \bR, \qquad c_v = \frac{1}{(\gamma-1)} \bR,
\end{equation}
such that the ratio of $c_p$ to $c_v$, called the isentropic constant $\gamma$, is equal to $5/3$.

It is well-known that the viscosity and temperature relation has a noticeable effect on the shock wave structure. Here we adopt the generally accepted temperature-dependent viscosity power law \citep{LC1992,Greenshields2007}: $\mu \propto T^s$ or $\mu = \alpha \, T^s$, where $\alpha$ is a constant of proportionality taken to be $\gamma^s$ and the power $s$ for almost all real gases falling between $0.5 \leq s \leq 1$, with the limiting cases, $s = 0.5$ and $s = 1$ corresponding to theoretical gases, namely, the hard-sphere and Maxwellian gases, respectively. In our simulations we use $s = 0.75$ for a monatomic argon gas. For a monatomic ideal gas the other transport coefficient, namely, the heat conductivity coefficient $\ka$ is related to the kinematic viscosity coefficient $\mu$ via the relation: $\ka = (c_p /\rm{Pr}) \mu$.

\subsection{\label{sec:III_A} Formulation of the the shock structure problem and numerical procedure}
We consider a planar shock wave propagating in the positive $x$-direction which is established in a flow of a monatomic gas. For this one-dimensional flow problem, all hydrodynamic variables are functions of a single spatial coordinate $x$ and time $t$ ; the system is assumed to be uniform (having no gradients) and infinite along the y- and z- directions.
The flow velocity and heat flux in the $x$-direction are denoted by $\mathit{u(x,t)}$ and $\mathit{q(x,t)}$, respectively, and are zero in the two remaining (y and z) orthogonal directions.  Further, it is straightforward to verify that the stress tensor has only one non-zero component, the longitudinal stress which can be expressed as
\begin{equation}
\label{eqn_S3}
\Pi_{\it xx} = -\,\frac{4}{3}\, \mu \, \frac{\partial u}{\partial x} \,-\,\frac{4}{3} \frac{\mu\,\kam}{\rho} \, \frac{\partial^2 \rho}{\partial x^2}\,+\,\frac{4}{3} \frac{\mu\,\kam}{\rho^2} \left( \frac{\partial \rho}{\partial x} \right)^2 \equiv \Pi,
\end{equation}
and the constitutive relation for the heat flux is
\begin{equation}
\label{eqn_S4}
q = - \ka \, \frac{\partial T}{\partial x} - \frac{c_p}{\rm{Pr}} \kam \,\rho \, T \,\frac{\partial \ln \rho}{\partial x}.
\end{equation}

With the above definitions, the one-dimensional reduced balance equations for the modified Navier-Stokes model can be written in `conservative' form,

\begin{align}
& \frac{\partial \rho}{\partial t} + \frac{\partial}{\partial x} \Big(\, \rho \, u \Big) = 0, \label{eqn_S5}\\
& \frac{\partial}{\partial t} \Big( \rho \, u \Big) + \frac{\partial}{\partial x} \Big(\, \rho \, u^2 + \rho \, \bR \, T\,  +\, \Pi\, \Big)  = 0,\label{eqn_S6}\\
& \frac{\partial}{\partial t} \left(\frac{1}{2}\rho\,  u^2\,  + \, C_v \, \rho \,T \right) + \frac{\partial}{\partial x} \left(\frac{1}{2}  \rho \, u^3 \, + C_p\, \rho \,  T \,u \, + \, \Pi \, u \, +\,  q \, \right) = 0. \label{eqn_S7}
\end{align}
The one-dimensional classical Navier-Stokes system is obtained by setting $\kam = 0$ in the constitutive relations of longitudinal stress and heat flux, i.e, in Eq. \eqref{eqn_S3} and Eq. \eqref{eqn_S4}, respectively. The corresponding Euler system is then obtained by setting $\Pi = 0$ and $q = 0$ in Eqs. \eqref{eqn_S5} - \eqref{eqn_S7}. A detailed dimensional analysis on the one-dimensional continuum flow model showing the importance of the corrections to the constitutive equations is included in the Appendix~\ref{AppendixA}.

The modified Navier-Stokes equations, for the one-dimensional stationary shock flow configuration reduced to:
\begin{eqnarray}
&& \dx \left[ \rho \, u \right] = 0,\label{eqn_S13}\\
&& \dx \left[ \rho\, u^2 \,+\,\rho \, \bR \, T \,+ \, \Pi\right] = 0, \label{eqn_S14}\\
&& \dx \Bigg[ \rho \, u \,\left(\frac{1}{2}\,u^2 \,+\,C_p\,T \right)\, + \, \Pi \,u \,+\,q \Bigg] = 0,\label{eqn_S15}
\end{eqnarray}
with the only non-zero longitudinal new shear stress $\Pi$ and the heat flux $q$ given by,
\begin{eqnarray}
&& \Pi = -\frac{4}{3} \mu \, \frac{{\rm d} u}{{\rm d} x}\,-\,\frac{4}{3} \frac{\mu\,\kam}{\rho} \, \frac{{\rm d}^2 \rho}{{\rm d} x^2}\,+\,\frac{4}{3} \frac{\mu\,\kam}{\rho^2} \left( \frac{{\rm d} \rho}{{\rm d} x} \right)^2, \label{eqn_S16}\\
&& q = - \ka \, \frac{{\rm d} T}{{\rm d} x} - \frac{c_p}{\rm{Pr}} \kam \,\, T \,\frac{{\rm d} \rho}{{\rm d} x}. \label{eqn_S17}
\end{eqnarray}

We denote the upstream ($x \to -\infty$) and downstream ($x \to \infty$) conditions of a shock, located at $x = 0$, by a subscript $1$ and $2$, respectively. That is the upstream and the down stream equilibrium states are characterised by  $(\rho_1, u_1, T_1)$ and $(\rho_2, u_2, T_2)$, respectively. Across a shock, the finite jump in each state variable is given by the so-called Rankine-Hugoniot (RH) relations/conditions~\citep{CF1948,LR1957} that connect the upstream and downstream states of a shock. RH relations provide necessary conditions for any solution of the system (\ref{eqn_S13} - \ref{eqn_S15}). The standard Rankine-Hugoniot relations for the one-dimensional stationary shock flow can be obtained from the conservation balance laws (Eqs. \eqref{eqn_S13} - \eqref{eqn_S15}) by following the standard procedure given in ~\citep{CF1948} and using the fact that the end states are in ``local'' equilibrium: there are no spatial variations in hydrodynamics fields which implies that the variables $\Pi$ and $q$ are zero at upstream and downstream end states, as:
\begin{align}
\rho_1 \,u_1 &= \rho_2 \,u_2 , \label{RH1}\\
\rho_{1} \,u_1^2 + \rho_{1} \,\bR \,T_{1} &= \rho_{2}\, u_2^2 + \rho_{2} \,\bR \,T_{2} , \label{RH2}\\
\rho_1\,u_1^3 + 2 \,c_p\, \rho_1\, T_1 \,u_1  &=  \rho_2\, u_2^3 + 2\, c_p\, \rho_2 \,T_2\, u_2. \label{RH3}
\end{align}
Integration of the system \eqref{eqn_S13} - \eqref{eqn_S15} leads to:
\begin{eqnarray}
&& \rho \, u = m_0,\label{eqn_S18}\\
&& \rho\,u \,+\, \rho\, \bR \,T\,+ \Pi = p_0, \label{eqn_S19}\\
&&  \rho \, u \left(c_p\, T + \frac{u^2}{2} \right) \,+ \,\Pi \,u\,+\, q  = m_0\,h_0,\label{eqn_S20}
\end{eqnarray}
where $m_0, p_0$ and $h_0$ are integration constants which represents the mass flow rate, the stagnation pressure and the stagnation specific enthalpy, respectively, and their values/expressions can be obtained using the well-known Rankine-Hugoniot conditions  \eqref{RH1} - \eqref{RH3}. In order to solve the system \eqref{eqn_S18} - \eqref{eqn_S20}, it is convenient to work with its dimensionless form. We use the following set of dimensionless variables based on the upstream reference states (denoted with subscript $1$) as in \citep{Reeseetal1995,RD2020,DR2020}:
\begin{align}
\begin{split}
\label{eqn_S21}
& \overline{\rho} = \frac{c_1^2}{p_1} \rho= \frac{\gamma}{\rho_1}\,\rho, \quad \overline{u} = \frac{u}{c_1}, \quad \overline{T} = \frac{\bR}{c_1^2} T, \quad \overline{p} = \frac{p}{p_1}, \quad  \overline{x} = \frac{x}{\la_1},\quad \overline{\mu} = \frac{\mu}{\mu_1},
\end{split}
\end{align}
where $\lambda_1$ is the upstream mean free path which is a natural choice for a characteristic length-scale as changes through the shock occur due to few collisions and $c_1 = \sqrt{\gamma\, \bR\, T_1}$ being the adiabatic sound speed. The upstream mean free path can be expressed as a function of reference state variables: $\lambda_1 = \lambda_0 \mu_1/\rho_1\,c_1$, with $\lambda_0 = (16/5)\,\sqrt{\gamma / 2\,\pi}$. Further, the dimensionless forms of transport coefficients $\overline{\ka}$ and $\overline{\ka}_m$ are:
\begin{equation}
\label{eqn_S22}
\overline{\ka} = \frac{\gamma}{(\gamma - 1)\,\rm{Pr}} \overline{\mu} \quad \text{and} \quad \overline{\ka}_m = \ka_{m_0} \frac{\overline{\mu}}{\overline{\rho}},
\end{equation}
with the Prandtl number, $\rm{Pr}$, is equal to $2/3$ for the case of a monatomic gas.

The nondimensionalized form of the integral conservation equations \eqref{eqn_S18} - \eqref{eqn_S20} can then be obtained using the dimensionless quantities defined via \eqref{eqn_S21} and \eqref{eqn_S22} as:

\begin{eqnarray}
\overline{\rho} \, \overline{u} &=& \overline{m}_0, \label{eqn_S23}\\
- \frac{1}{\la_0\,\rm{Ma}_1} \, \overline{\Pi} &=& \frac{\overline{T}}{\overline{u}}\,+\,\overline{u}\,-\overline{p}_0, \label{eqn_S24}\\
- \frac{\left(\gamma - 1\right)}{\la_0\,\rm{Ma}_1} \, \overline{q} &=& \overline{T}\,-\,\frac{\left(\gamma - 1\right)}{2} \overline{u}^2\,+\,\left(\gamma - 1 \right) \,\overline{p}_0\, \overline{u}\,-\,\overline{h}_0, \label{eqn_S25}
\end{eqnarray}

where $\rm{Ma}_1$ is the upstream Mach number defined as the ratio of the speed of the gas to the speed of sound through the gas, ${\rm{Ma}_1} = u_1/c_1$. Expressions for the quantities $\overline{m}_0$, $\overline{p}_0$ and $\overline{h}_0$ can be then obtained as
\begin{equation}
\label{eqn_S26}
\overline{m}_0 = \gamma \, \rm{Ma}_1, \quad \overline{p}_0 = \frac{1}{\gamma\, \rm{Ma}_1} \, \left(1\,+\, \gamma\, \rm{Ma}_1^2 \right), \quad \overline{h}_0 = 1\,+\,\frac{\left(\gamma\,-1 \right)}{2} \rm{Ma}_1^2,
\end{equation}
and the expressions for the dimensionless shear stress ($\overline{\Pi}$) and the heat flux ($\overline{q}$) are given by,

\begin{eqnarray}
&& \overline{\Pi} = -\frac{4}{3} \overline{\mu} \, \frac{{\rm d} \overline{u}}{{\rm d} \overline{x}}\,-\,\frac{4}{3} \left( \frac{\gamma}{\la_0}\right) \frac{\overline{\mu}\,\overline{\ka}_m}{\overline{\rho}} \, \frac{{\rm d}^2 \overline{\rho}}{{\rm d} \overline{x}^2}\,+\,\frac{4}{3} \left(\frac{\gamma}{\la_0}\right) \frac{\overline{\mu}\,\overline{\ka}_m}{\overline{\rho}^2} \left( \frac{{\rm d} \overline{\rho}}{{\rm d} \overline{x}} \right)^2,\label{eqn_S27}\\
&& \overline{q} = -\overline{\ka} \, \frac{{\rm d} \overline{T}}{{\rm d} \overline{x}} \,-\,\frac{\gamma}{\left(\gamma - 1\right)\, \rm{Pr}}\,  \overline{\ka}_m\, \overline{T}\, \frac{{\rm d} \overline{\rho}}{{\rm d} \overline{x}}.\label{eqn_S28}
\end{eqnarray}

We solve the final system \eqref{eqn_S23}--\eqref{eqn_S25} using a numerical scheme, namely, finite difference global solution (FDGS)  developed by \citet{Reeseetal1995} with well-posed boundary conditions. The specific details of FDGS scheme can be found in \citep{Reeseetal1995}.

\section{\label{sec:IV} Macroscopic fields across the shock layer}
\begin{figure}
\includegraphics[width=7.5cm]{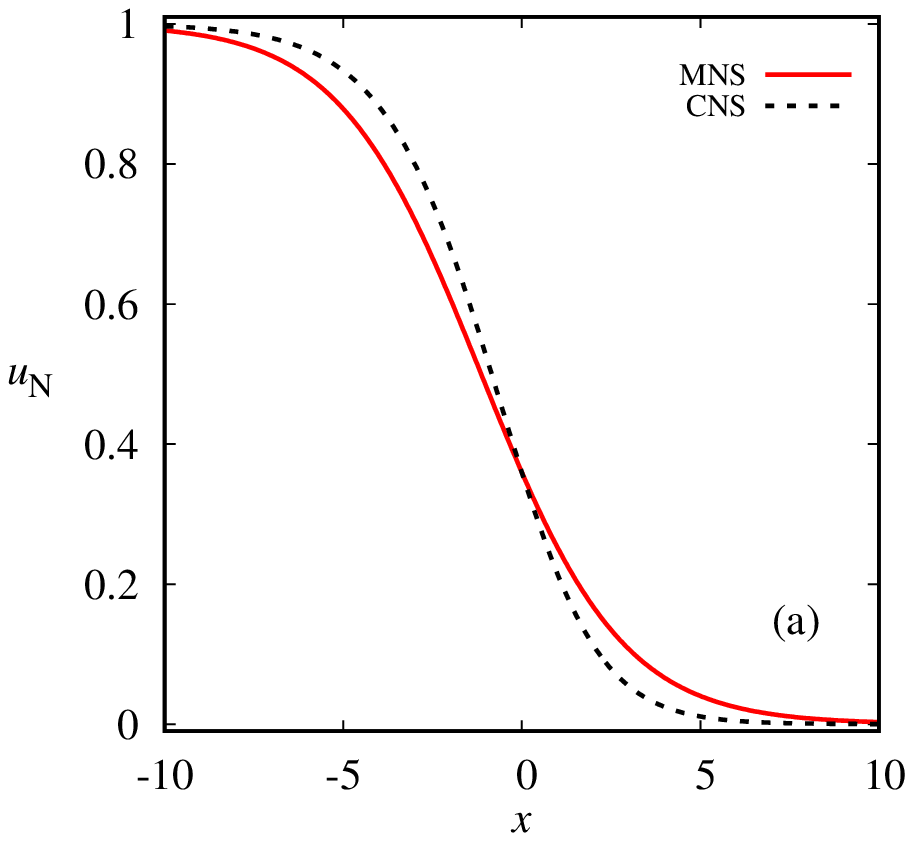}\,
\includegraphics[width=7.5cm]{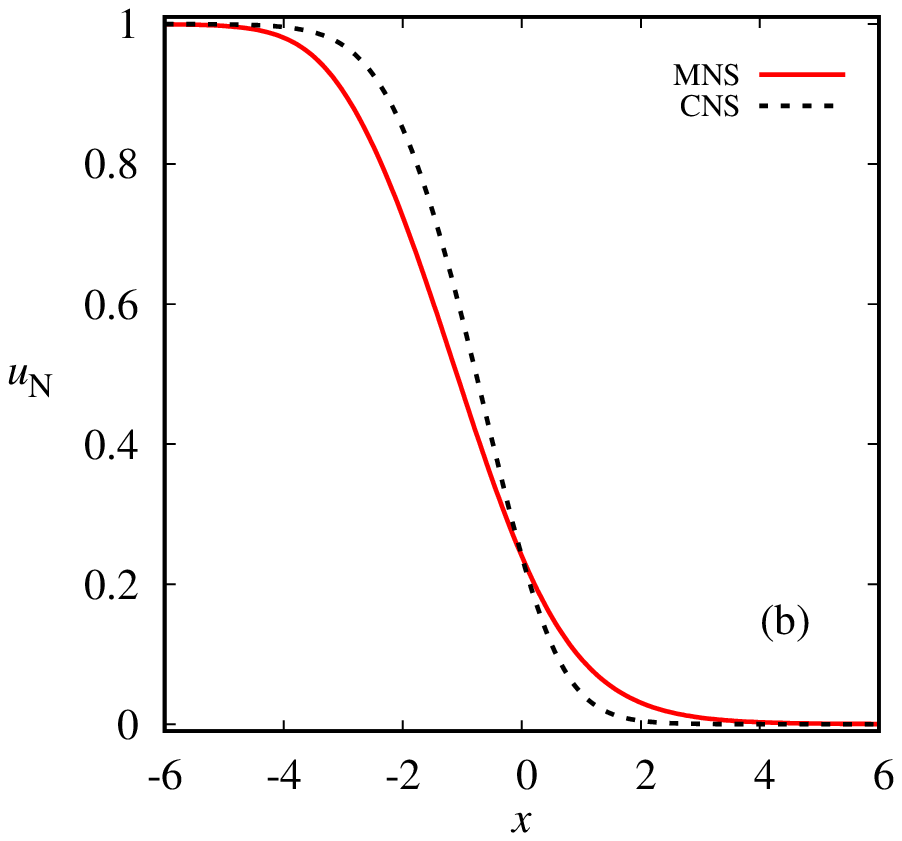}
\caption{Variation of normalized velocity ($u_{\rm N}$) profiles in $\rm{Ar}$ shock layer: for (a) $\rm{Ma}_1 = 1.55$, (b) $\rm{Ma}_1 = 3.38$.}
\label{fig:1}
\end{figure}

We perform numerical simulations of stationary shock waves located at $x=0$ using FDGS scheme by considering a computational spatial domain of length $50 \lambda_1$ covering $(-25 \lambda_1, 25 \lambda_1)$ with 1000 spatial grid points. This is wide enough to contain the entire shock profile ranging from supersonic to hypersonic without altering its structure. The constant $\ka_{m_0}$ in the molecular mass diffusivity coefficient $\ka_m$ is set to $0.5$ \cite{RD2021} in all the present simulations. To compare the shock structure profiles among the theoretical and experimental data, the position $x$ has been scaled such that $x = 0$ corresponds to a value of the normalized gas density $\rho_{\rm N} = (\rho - \rho_1)/(\rho_2 - \rho_1)$ equals $0.5$. Other hydrodynamic fields, namely, the velocity and the temperature profiles are normalised via: $ u_{\rm N} = (u - u_2) / (u_1 - u_2),$ and $\theta_{\rm N} = (\theta - \theta_1) / (\theta_2 -\theta_1),$ respectively. Figure \ref{fig:1} (a) and (b) shows the comparison of the normalised velocity profiles obtained from the classical and the new modified Navier-Stokes equations for $\rm{Ma}_1 = 1.55$ and $\rm{Ma}_1 = 3.38$, respectively. As the flow varies from hypersonic/supersonic to subsonic across the shock, the velocity is maximum/high at the upstream part of the shock, decreases through the shock and  attains its smallest value at the downstream part of the shock. The velocity profiles obtained from the modified NS model are more diffusive than the classical NS profile at both upstream and downstream part of the shock which is evident from Fig. \ref{fig:1}. However, for hypersonic flow for which $\rm{Ma}_1 \geq 6.5$ (figure not shown) we observed that the modified NS profiles are steepened at the upstream part of the shock but still more diffusive than the classical profiles.

\begin{figure}
\includegraphics[width=7.5cm]{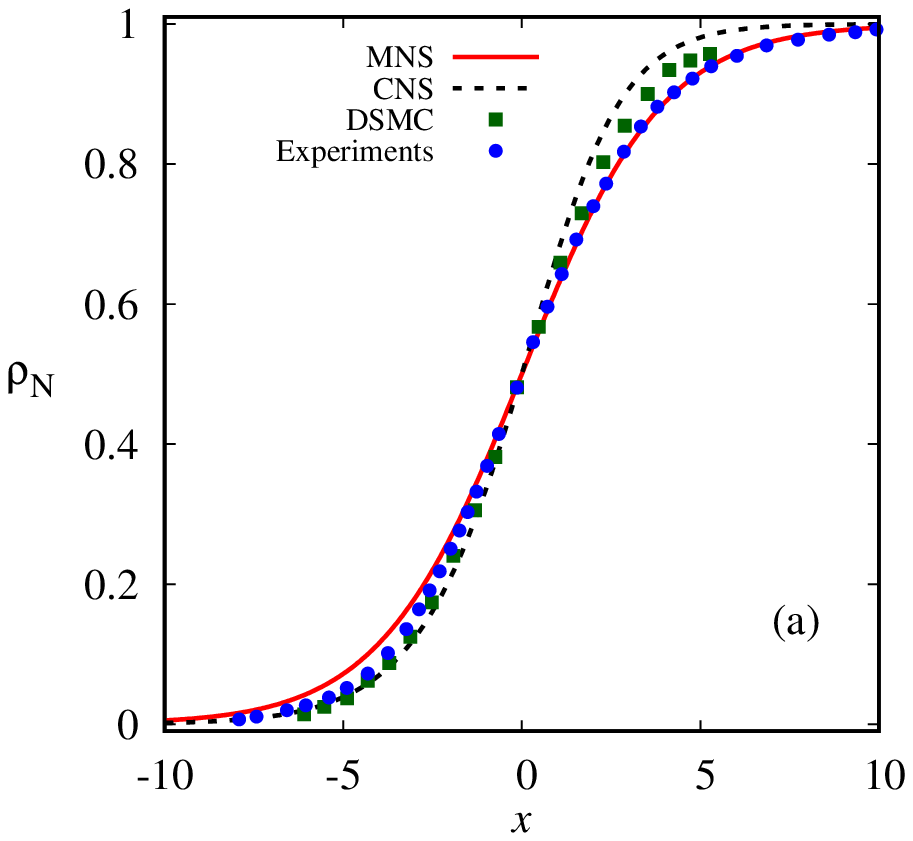}\,
\includegraphics[width=7.5cm]{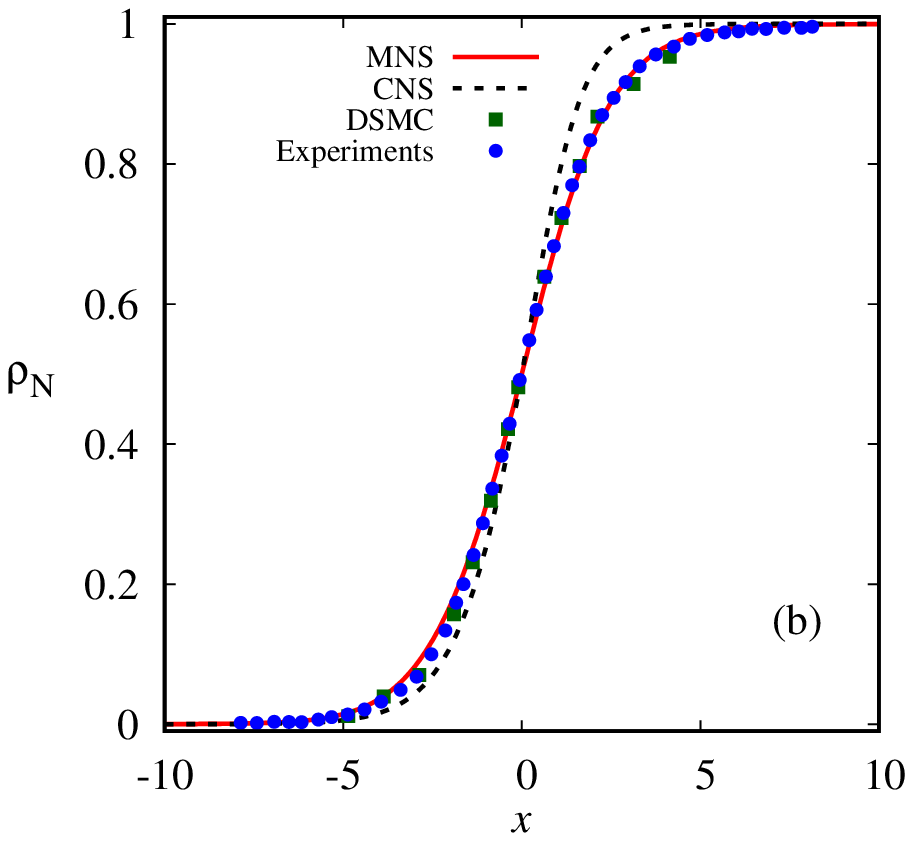}\\
\includegraphics[width=7.5cm]{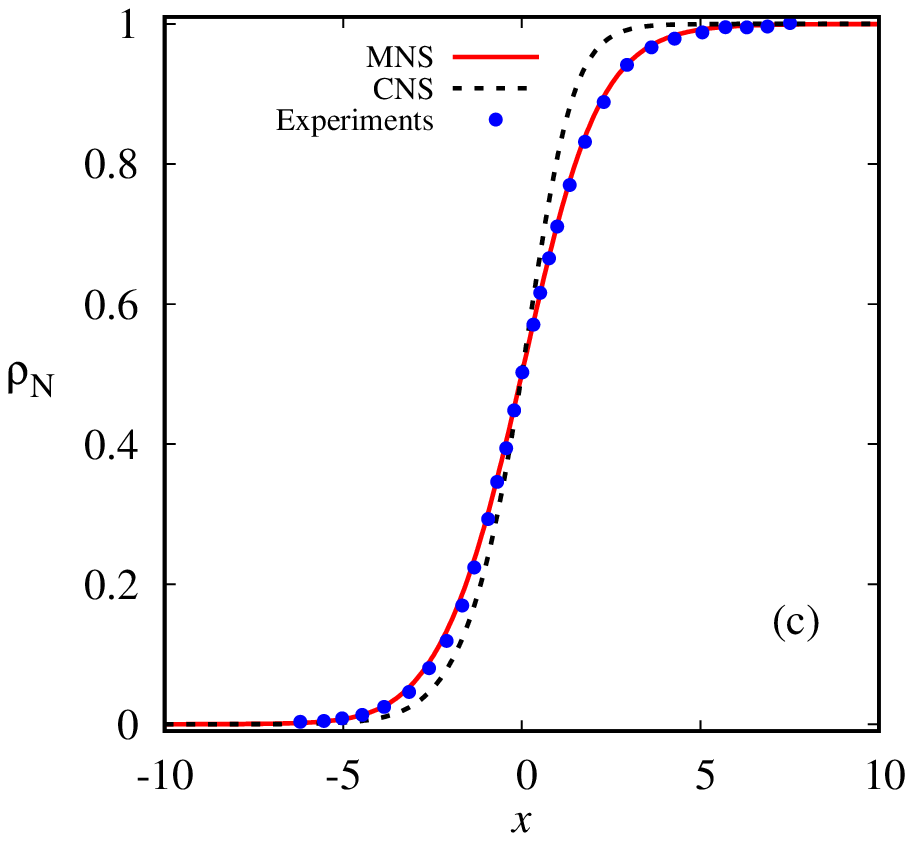}\,
\includegraphics[width=7.5cm]{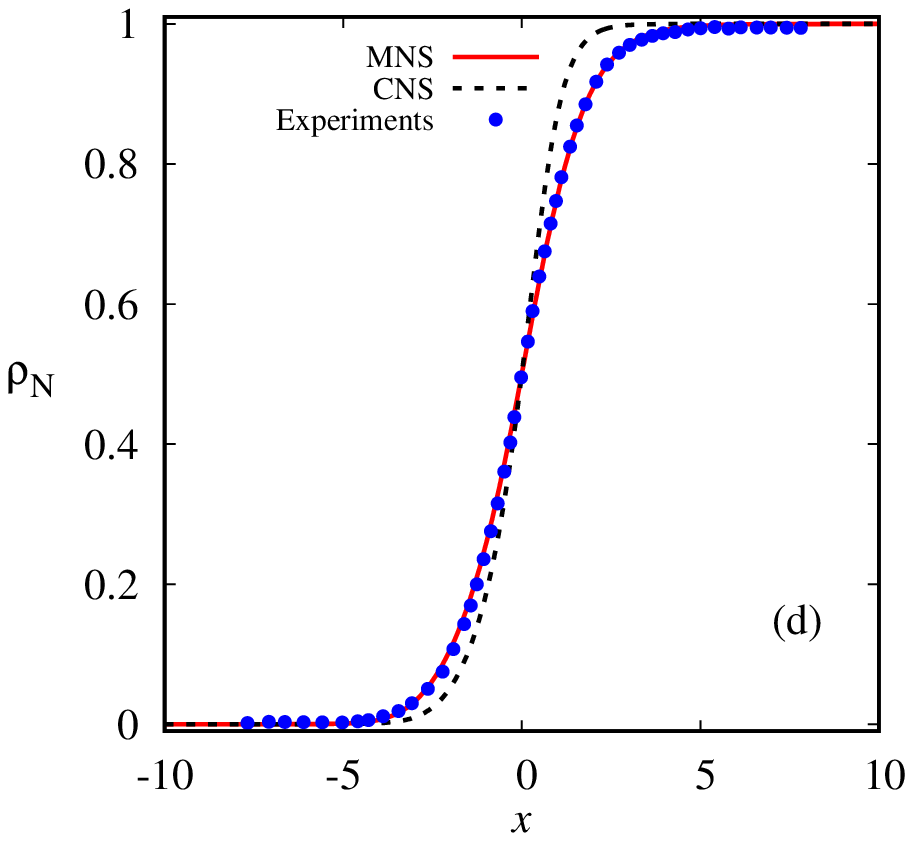}
\caption{Variation of normalized density ($\rho_{\rm N}$) profiles in $\rm{Ar}$ shock layer: for (a) $\rm{Ma}_1 = 1.55$, (b) $\rm{Ma}_1 = 2.05$, (c) $\rm{Ma}_1 = 2.31$ and (d) $\rm{Ma}_1 = 3.38$. In each panel, the dashed line (black) represent the solution of classical Navier-Stokes equations (CNS), the solid line (red) represent the solution of modified Navier-Stokes equations (MNS) and the filled circles (blue) represent the experimental data of \citet{Alsmeyer1976}. The DSMC data from \citet{Alsmeyer1976} is superimposed in panels (a) and (b) as filled squares (green).}
\label{fig:2}
\end{figure}

\subsection{\label{sec:IV_A} Density profiles}
\begin{figure}
\includegraphics[width=7.5cm]{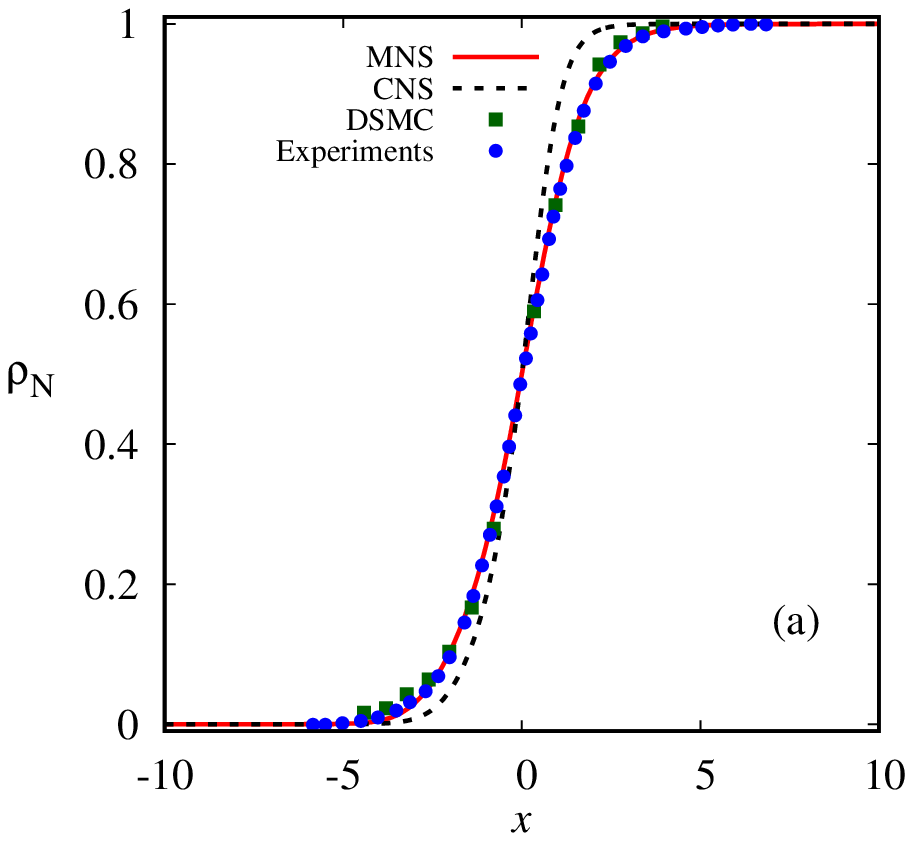}\,
\includegraphics[width=7.5cm]{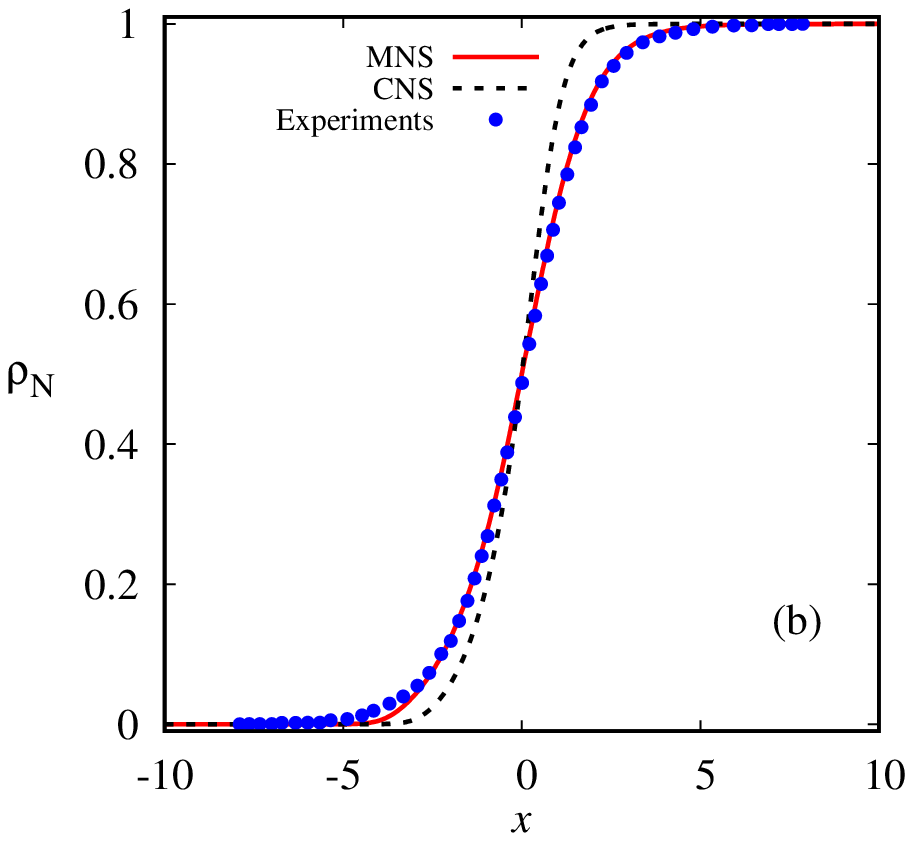}\\
\includegraphics[width=7.5cm]{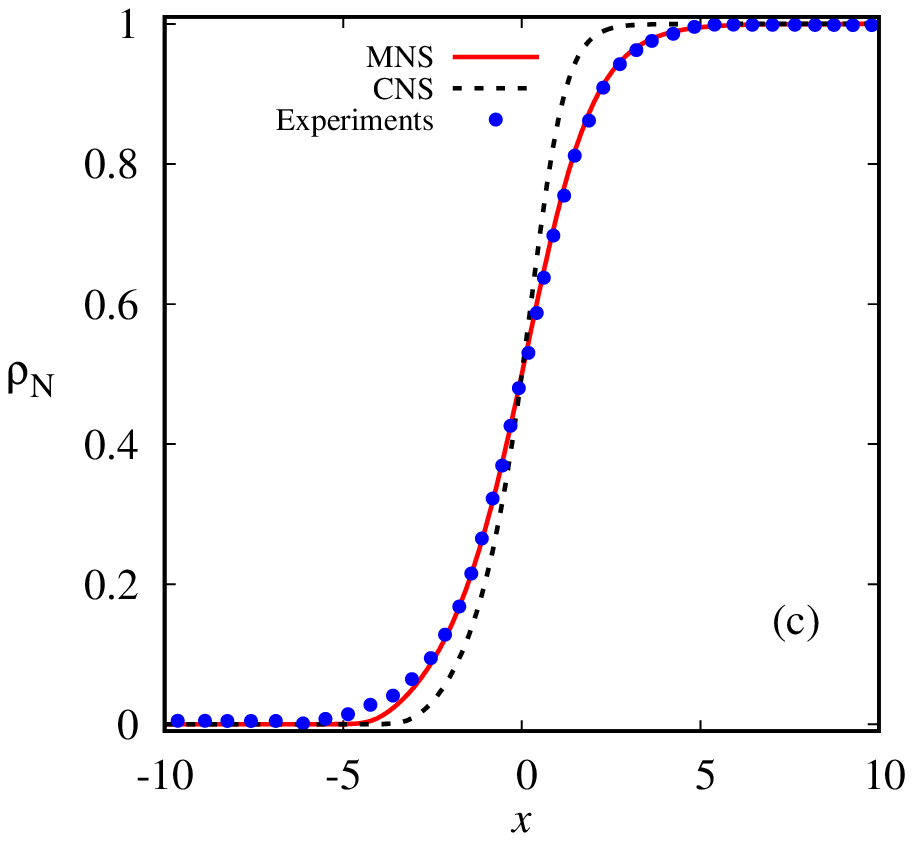}\,
\includegraphics[width=7.5cm]{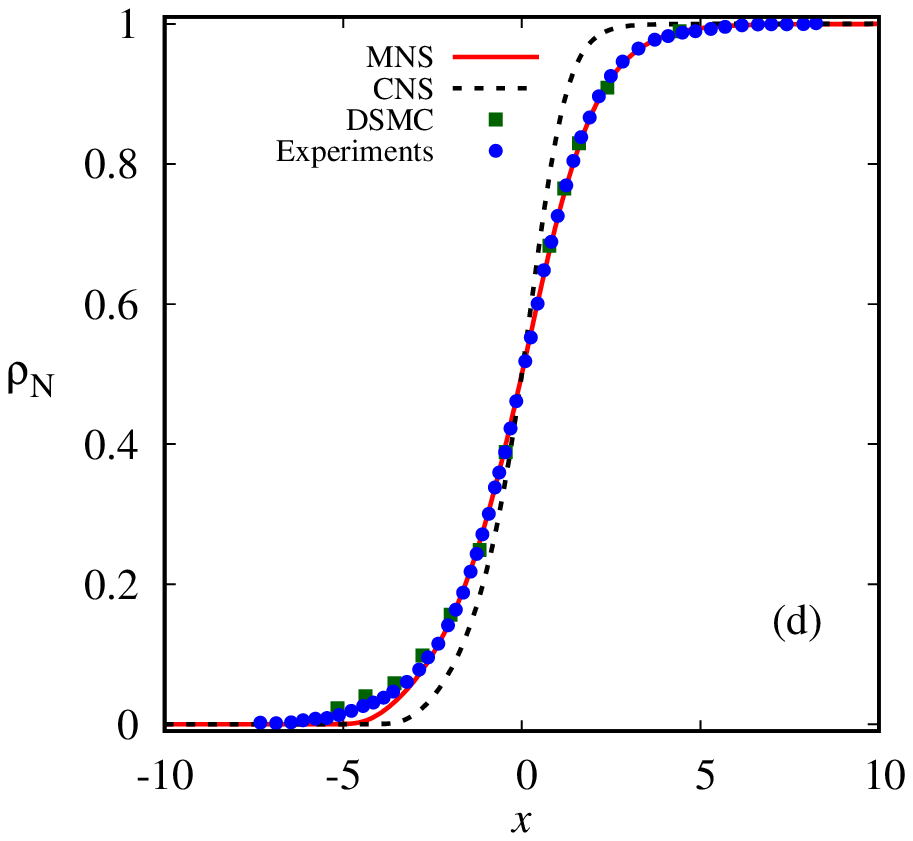}
\caption{Variation of normalized density ($\rho_{\rm N}$) profiles in $\rm{Ar}$ shock layer: for (a) $\rm{Ma}_1 = 3.8$, (b) $\rm{Ma}_1 = 6.5$, (c) $\rm{Ma}_1 = 8$ and (d) $\rm{Ma}_1 = 9$.  In each panel, the dashed line (black) represent the solution of classical Navier-Stokes equations (CNS), the solid line (red) represent the solution of modified Navier-Stokes equations (MNS), the filled circles (blue) represent the experimental data of \citet{Alsmeyer1976}. The DSMC data from \citet{Alsmeyer1976} is superimposed in panels (a) and (d) as filled squares (green).}
\label{fig:3}
\end{figure}

Experimental data exist for monatomic argon gas density variations across shock layer \citep{Alsmeyer1976}. We take a detailed comparison of density field across the shock with experimental results of \citet{Alsmeyer1976} and also with the classical Navier-Stokes equations. Figure \ref{fig:2} shows the normalized density $\rho_N$ profiles through an argon shock wave as predicted by the modified Navier-Stokes and the classical Navier-Stokes equations compared with the experimentally measured density data and available Bird’s Monte Carlo simulation (later called DSMC - direct simulation Monte Carlo) data from \citet{Alsmeyer1976}. Panels (a), (b), (c) and (d) of Fig. \ref{fig:2} correspond to upstream Mach numbers of $\rm{Ma}_1 = 1.55, 2.05, 2.31$ and $3.38$, respectively. In each panel: the dashed line (black) and the solid line (red) represent solutions of the Navier-Stokes equations and the modified Navier-Stokes equations, respectively. The filled circles (blue) represent the experimental data of \citet{Alsmeyer1976} and the filled squares (green) superimposed in panels (a) and (b) represent the DSMC data. From panel (a) of Fig. \ref{fig:2} one observes that for the upstream Mach number of $1.55$ the classical Navier-Stokes is able to predict well the upstream part of the shock layer in comparison with the experimental data but completely fails to predicts the downstream part of the shock layer. While the modified Navier-Stokes equations produce good agreement with the experimental data with a small disparity at the upstream part of the shock layer and is more diffusive than the experimental data. Further the modified Navier-Stokes model produce good agreement with the DSMC data in the middle of the shock and is more diffusive at both the upstream and the downstream part of the shock as it is evident from panel (a) of Fig. \ref{fig:2}. The modified Navier-Stokes predictions for the normalized density profiles show excellent agreement with the experimental data for the upstream Mach number of $\rm{Ma}_1= 2.05, 2.31, 3.38$ and $3.8$, which is evident from panels (b)-(d) of Fig. \ref{fig:2} and panel (a) of Fig. \ref{fig:3}, respectively, and also show excellent agreement with the DSMC data for the upstream Mach number of $\rm{Ma}_1= 2.05$ and $\rm{Ma}_1= 3.8$, which is visible from panel (b) of Fig. \ref{fig:2} and panel (a) of Fig. \ref{fig:3}, respectively. Overall, an excellent agreement between predictions of the modified Navier-Stokes equations and the experimental data of \citet{Alsmeyer1976} is found for weak shocks ($\rm{Ma}_1 \sim 1$) to moderate strong shocks ($\rm{Ma}_1 \sim 3$).

Panels (b), (c) and (d) of Fig.~\ref{fig:3} show the normalized density profiles comparison between the modified and the classical Navier-Stokes equations along with the experimental results for hypersonic upstream Mach numbers of $\rm{Ma}_1 = 6.5, 8$ and $9$, respectively. It is seen in Fig.~\ref{fig:3} (b), (c) and (d) that the predictions by modified Navier-Stokes equations for the variation of the density within the shock layer are in very good agreement with the experimental data, however, one can notice that the modified NS profiles are slightly less diffusive compared to the experimental profiles and are more diffusive than the classical profiles at the upstream part of the shock layer. Similar kind of behaviour is reported in \citet{Paolucci2018} using a second-order continuum theory for density profiles in argon gas at $\rm{Ma}_1 = 9$. Overall, from Fig.~\ref{fig:2} and Fig.~\ref{fig:3} we conclude that the modified Navier-Stokes solutions are in excellent agreement with the experimental results of \citet{Alsmeyer1976}; are in good agreement with the  Bird’s Monte Carlo simulation (DSMC) data; and are better than the classical Navier-Stokes prediction at all upstream Mach numbers discussed here.

\subsection{\label{sec:IV_B} Temperature profiles}

\begin{figure}
\includegraphics[width=7.5cm]{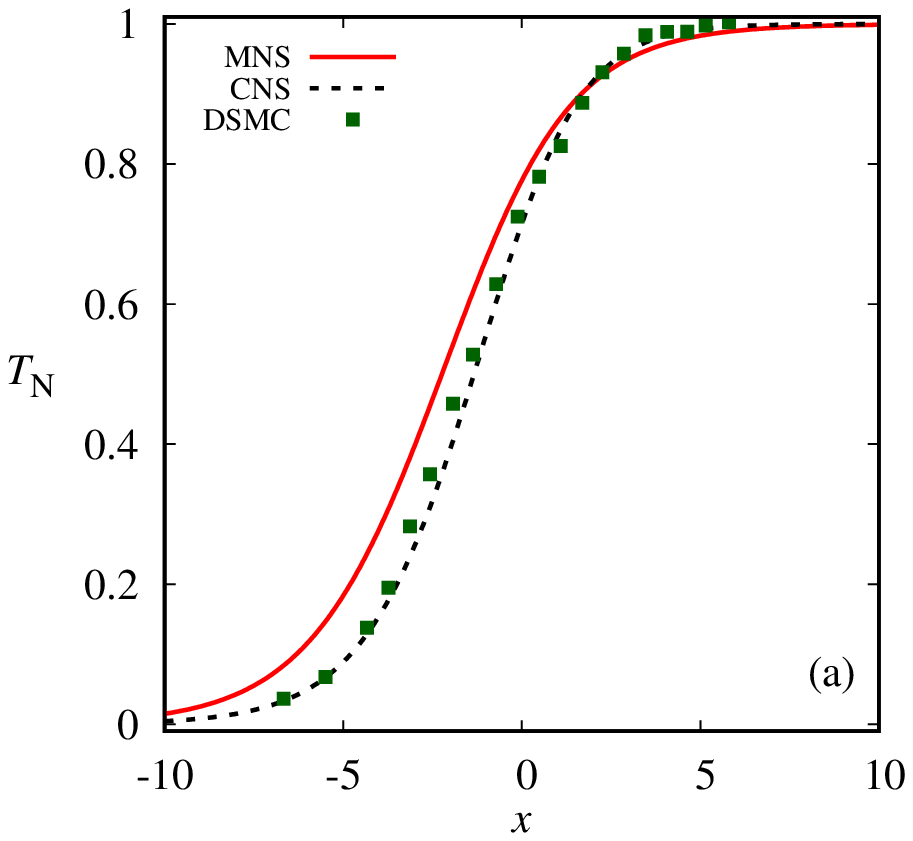}\,
\includegraphics[width=7.5cm]{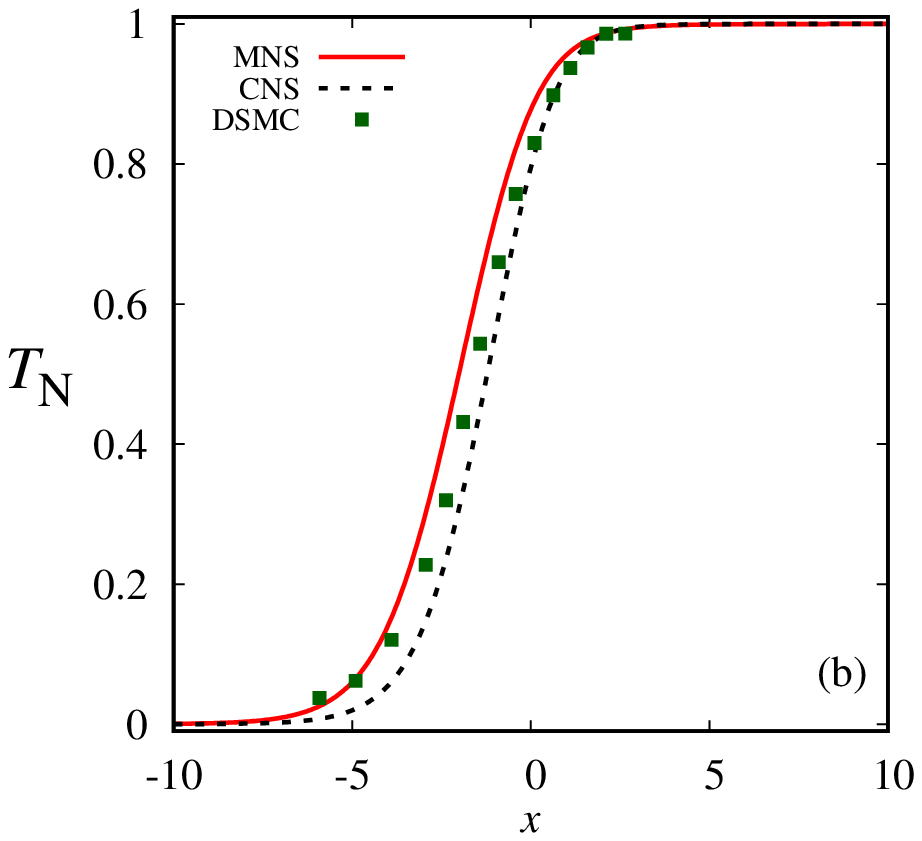}\\
\includegraphics[width=7.5cm]{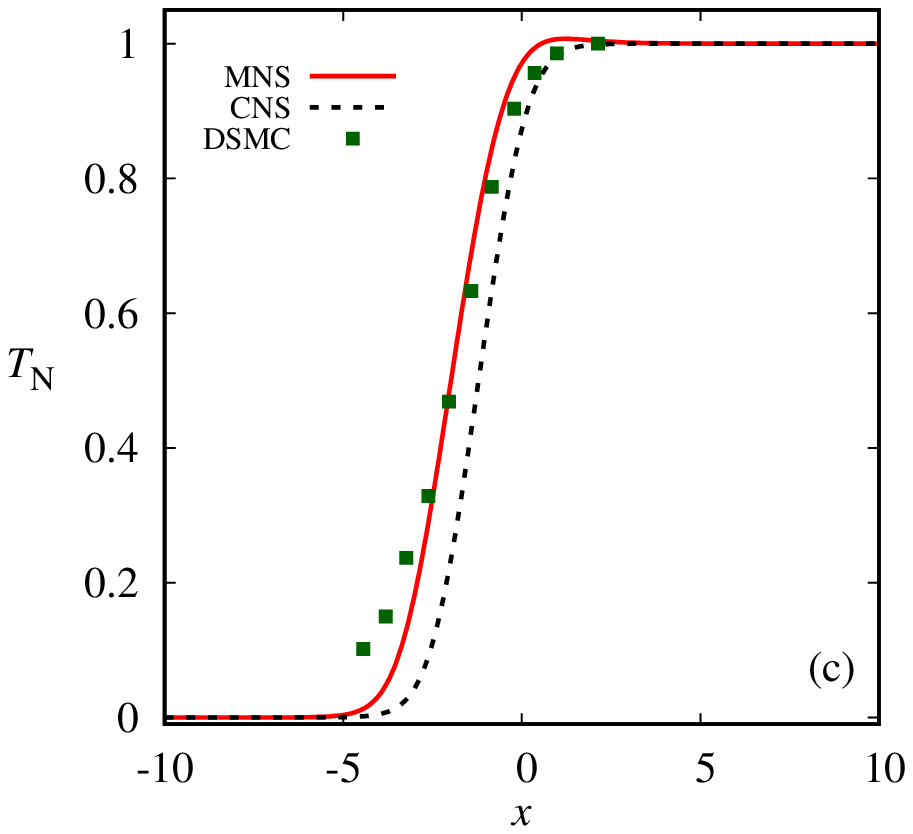}\,
\includegraphics[width=7.5cm]{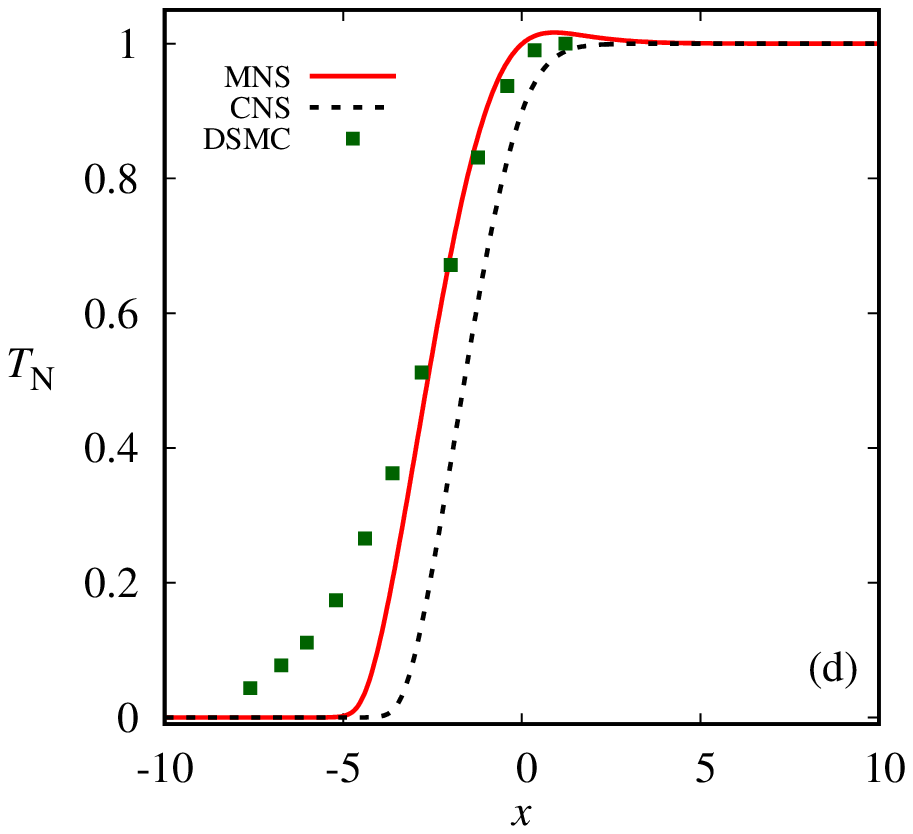}
\caption{Variation of normalized temperature ($T_N$) profiles in $\rm{Ar}$ shock layer: for (a) $\rm{Ma}_1 = 1.55$, (b) $\rm{Ma}_1 = 2.05$, (c) $\rm{Ma}_1 = 3.8$ and (d) $\rm{Ma}_1 = 9$. In each panel, the dashed line (black) represent the solution of classical Navier-Stokes equations (CNS), the solid line (red) represent the solution of modified Navier-Stokes equations (MNS) and the filled squares (green) represent the DSMC data taken from \citet{Alsmeyer1976}.}
\label{fig:4}
\end{figure}

Due to unavailability of experimental data for the temperature profiles within the shock layer here we make use of available temperature profiles from Bird's Monte Carlo simulation
data (DSMC data) with $s = 0.72$ which is reported in \citet{Alsmeyer1976} and are limited for only few upstream Mach numbers.
Comparison of shock temperature profiles are rarely reported due to the fact that the temperature is more sensitive quantity since it is a higher-order moment of the velocity distribution. No clear comparison of the prediction of the temperature profiles by the major theoretical models including Grad's 13-moment model, Regularized 13-moment equations, Burnett equations are reported in the literature. However, here in Fig.~\ref{fig:4}, we report the comparison of normalized temperature $T_{\rm N}$ profiles as predicted by the modified Navier-Stokes equations and the classical Navier-Stokes equations with the Bird's DSMC results assembled in \citet{Alsmeyer1976}. Panels (a), (b), (c) and (d) of Fig.~\ref{fig:4} correspond to upstream Mach numbers of $\rm{Ma}_1 = 1.55, 2.05, 3.8$ and $9$, respectively, with the dotted black lines representing the solutions of the classical Navier-Stokes, the solid red lines representing the solutions by the modified Navier-Stokes equations and the filled blue circles represent the Bird's Monte Carlo simulation data from \citet{Alsmeyer1976}. From Fig.~\ref{fig:4} one can observe that predictions by the modified Navier-Stokes equations are in qualitative agreement with the Monte Carlo simulation data at high upstream Mach numbers ($\rm{Ma}_1 > 3$).

\begin{figure}
\centering
\includegraphics[scale=1.0]{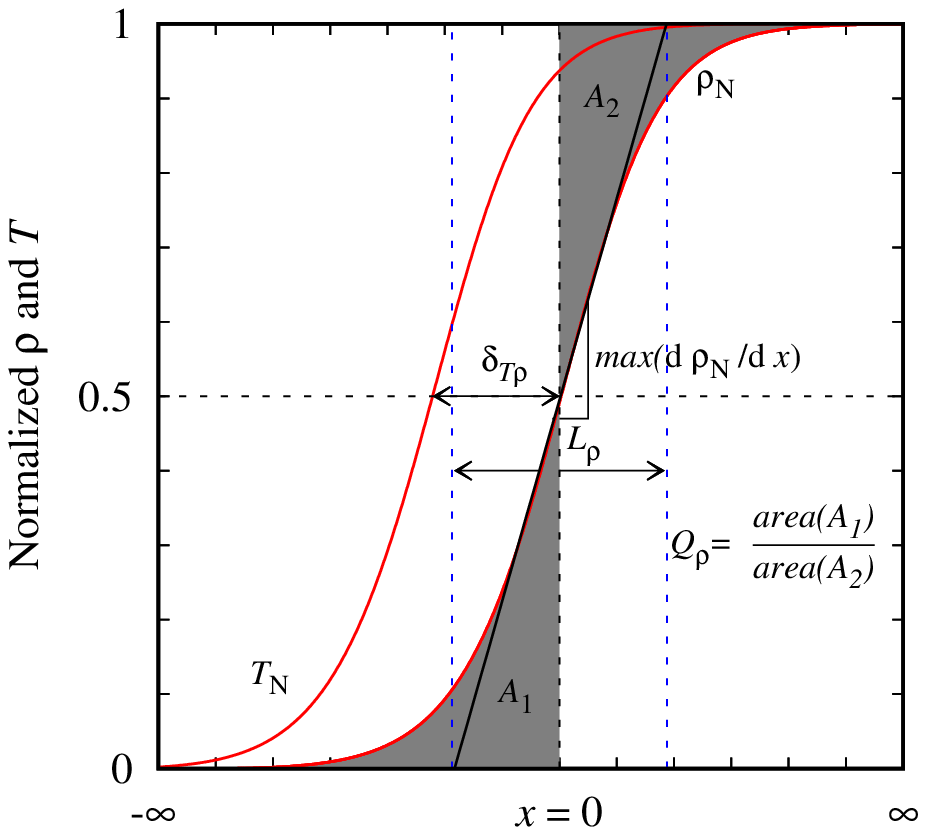}
\caption{Schematic of shock macroscopic parameters - shock thickness $(L_{\rho} )$, shock asymmetry $(Q_{\rho})$ and temperature-density separation $(\delta_{T\rho})$.}
\label{fig:5}
\end{figure}

\subsection{\label{sec:IV_C} Shock macroscopic parameters: shock thickness, density asymmetry and temperature-density separation}
In this section, we discuss three important parameters, namely, shock thickness, $L_{\rho}$, density asymmetry factor, $Q_{\rho}$, and temperature-density separation, $\delta_{T\rho}$, which are often used to characterize shock wave properties instead of comparing the full shock wave profile. Schematic of these three shock parameters are illustrated in Fig.~\ref{fig:5}. Frequently, in studying shock structures only the first two shock parameters (shock thickness and density asymmetry) are validated against experimental data (where available) and other numerical simulation results. These two shock macroscopic parameters are defined based on shock density profiles as seen from schematic diagram in Fig.~\ref{fig:5}.

The usual shock thickness or width, $L_{\rho}$, is defined as \cite{GP1953,Alsmeyer1976,PEM1991,Greenshields2007}:
\begin{equation}
\label{eqn_SW}
L_{\rho} = \frac{\rho_2 - \rho_1}{|\max(\frac{{\rm d} \rho}{{\rm d} x})|},
\end{equation}
and is based on the density profile and depends mainly on the central part of the shock wave. Note that from the definition of $L_{\rho}$, one can infer that it has a linear dependence on the density difference between the upstream and downstream states and inversely proportional to a slope corresponding to the maximum density gradient. In general, the non-dimensional inverse shock thickness, $\delta = \la_1/L_{\rho}$, is used instead of shock thickness, $L_{\rho}$, to compare computational results with experiments as it possesses an important feature that is, it represents actually the Knudsen number of the shock structure flow problem. In other words, one can say that the shock thickness, $L_{\rho}$, acts as the characteristic dimension of the flow configuration \cite{Greenshields2007}.

\begin{figure}
\includegraphics[width=8.5cm]{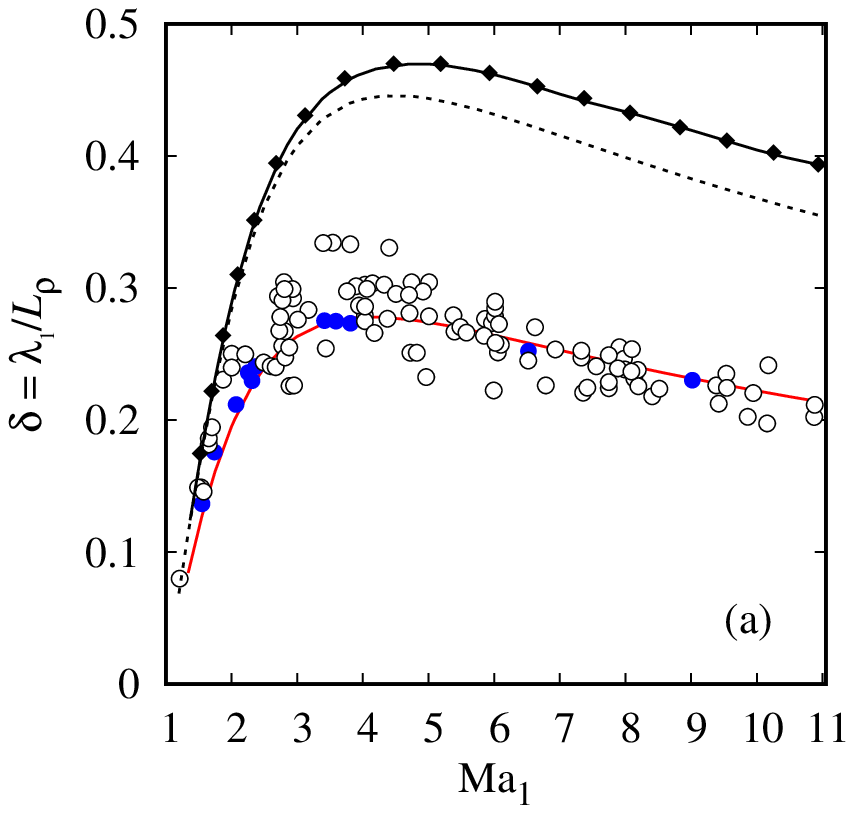}\,
\includegraphics[width=8.5cm]{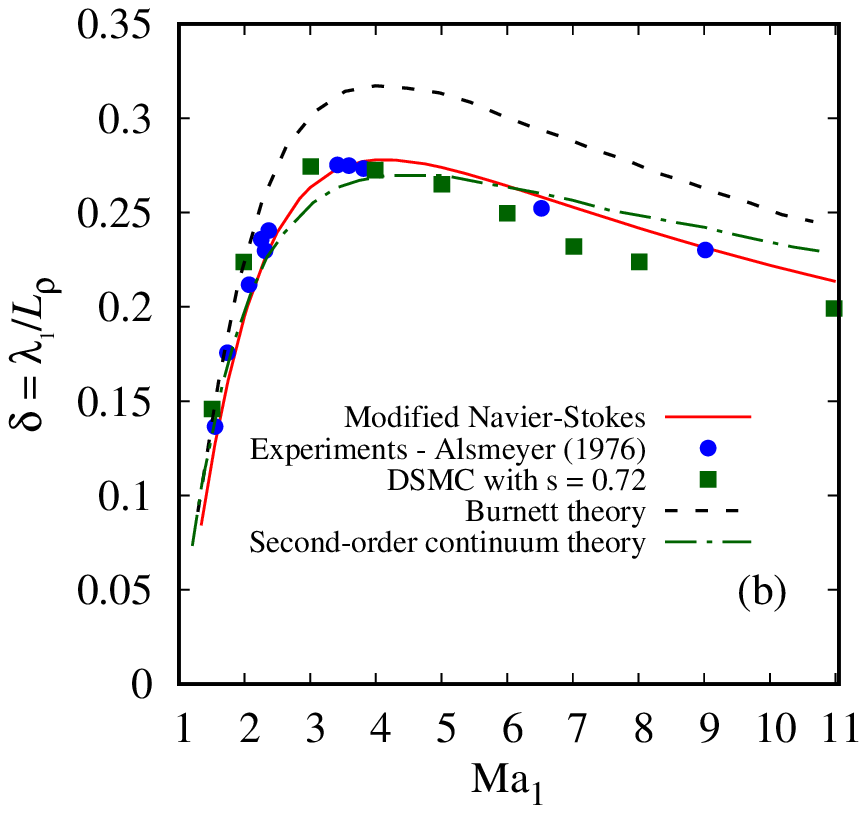}
\caption{(a,b) Variation of the reciprocal shock thickness, $\delta$, as a function of upstream Mach number, $\rm{Ma}_1$, for argon gas. Details on the description of plotted data is indicated in Table~\ref{table:data}.}
\label{fig:6}
\end{figure}

The most extensive collection of experimental data for the reciprocal shock thickness ($\delta$) in argon gas is recorded in \cite{Alsmeyer1976}. In Fig.~\ref{fig:6} the reciprocal shock thickness as a function of upstream Mach number predicted by the modified Navier-Stokes, the classical Navier-Stokes and other theories such as Burnett and a second-order continuum theories is compared with experimental data (open and filled circles) and also with the DSMC simulation data (filled squares). To access the accuracy of the numerical scheme (FDGS technique), the predictions by the classical Navier-Stokes equations with $s = 0.72$ (black solid line) using the FDGS technique are compared with the predictions of NS model with $s = 0.72$ (filled diamond symbols) using other numerical scheme presented in \citet{Paolucci2018}. Numerical predictions by classical Navier-Stokes using FDGS technique shows excellent agreement with the numerical results of \citet{Paolucci2018} (see panel (a) of Fig.~\ref{fig:6}) and this validates the accuracy of numerical scheme used here. Figure \ref{fig:6}(a) further shows the predictions of the modified Navier-Stokes equations for the reciprocal shock thickness (the inverse density thickness) in argon gas for an upstream Mach number up to $\rm{Ma}_1 = 11$, with experimental data assembled from \citet{Alsmeyer1976}. Predictions from the classical Navier-Stokes with $s = 0.75$ (see black short dashed line) are also presented for the sake of completeness. From Fig. \ref{fig:6} (a), one can observe that the classical Navier-Stokes equations with $s = 0.75$  and with $s = 0.72$ (see black solid line) predict the reciprocal shock thickness to be $1.4$ to $2$ times the measured values over the entire Mach number range presented. In other words we can say that the classical Navier-Stokes equations predict very small values for the shock thickness ($L_{\rho}$). Predictions from the modified Navier-Stokes equations with $s = 0.75$ is found to follow very closely the experimental results of \citet{Alsmeyer1976} (see filled blue circles) and is also in good agreement with the other experimental data.

\begin{table}[htbp]
\caption {Description of the plotted data in Fig.~\ref{fig:6} and Fig.~\ref{fig:7}} \label{table:data}
\begin{ruledtabular}
\begin{tabular}{cccc}
  Data label (line/symbol) & Description of the data \\
  \hline
  $\protect \redline$    &  modified NS solution with $\ka_{m_0} = 0.5$ and $s = 0.75$\\  
  $\protect \dashed$    & classical NS solution with $s = 0.75$ using FDGS technique\\ 
  $\protect \blackline$    & classical NS solution with $s = 0.72$ using FDGS technique\\ 
  $\protect \filleddaimond$    & classical NS solution with $s = 0.72$ from \citet{Paolucci2018}\\ 
  $\protect \bluecircle$    & experimental results of \citet{Alsmeyer1976} \\ 
  $\protect \opencircle$    & experimental results of \citet{LH1963}, \citet{Camac1965},\\
  & \citet{Sf1965}, \citet{Russell1965}, \citet{RT1966}, \\
  & \citet{Schmidt1969}, , \citet{Rieutord1970} and  \citet{Garenetal1974}. \\
  $\protect \filledsquare$  & DSMC results with $s = 0.72$ taken from \citet{LC1992} \\
  $\protect \longdashed$    & Burnett theory results of \citet{LC1992}\\ 
  $\color{darkgreen} \protect \dashdot$    & second-order continuum theory results of \citet{Paolucci2018}\\ 
  $\protect \dashdotdot$   & Brenner Navier-Stokes results of \citet{Greenshields2007}\\
  $\protect \filledtriangle$   & recast Navier-Stokes results of \citet{RD2020}\\
\end{tabular}
\end{ruledtabular} 
\end{table}

In panel (b) of Fig.~\ref{fig:6}, the predictions of the modified Navier-Stokes equations for the reciprocal shock thickness as a function of upstream Mach number are compared with the theoretical results from Burnett equations \citep{LC1992} (black long dashed line) and second-order continuum equations of \citet{Paolucci2018} (green dash dot line) along with the Alsmeyer experimental data (filled blue circles) and also with the DSMC data (filled green squares). It can be seen that the prediction by the modified NS equations shows an excellent agreement with the Alsmeyer experimental results and a close reasonable agreement with the DSMC results of \citet{LC1992} at all upstream Mach numbers ranging from $\rm{Ma}_1= 1.55$ to $9$. Some visible deviations from the Alsmeyer experimental results are present in Burnett (see Fig.~\ref{fig:6} (b)). Overall, judged by the inverse shock thickness, it is found from Fig.~\ref{fig:6} that the modified Navier-Stokes model gives good agreement with the experimental results of \citet{Alsmeyer1976} and a reasonable good agreement with the DSMC results \cite{LC1992}. 

Shock thickness, $L_{\rho}$, does not express anything about the overall shape of the shock due to the fact that it depends only on the maximum density gradient around the middle of the shock. In fact, shock thickness fails to provide adequate detailed information about shock profile: i.e., it may be possible that the experimentally measured profiles and calculated profiles from different theories may differ considerably with the shock thickness being the same. Therefore, a second important shock macroscopic parameter called the density asymmetry factor, $Q_{\rho}$, can be used to describe the actual shape of the shock structure as it measures skewness of the density profile relative to its midpoint \cite{Greenshields2007}. The shock asymmetry, $Q_{\rho}$, is defined based on the normalized density profile, $\rho_{\rm N}$, with its centre, $\rho_{\rm N} = 0.5$, located at $x = 0$, as
\begin{equation}
\label{eqn_Asym}
Q_{\rho} = \frac{\int_{-\infty}^{0} \rho_{N}(x)\, dx}{\int_{0}^{\infty} \left[ 1\,-\,\rho_N(x)\right] dx} \equiv \frac{\text{area of region } A1}{\text{area of region } A2}.
\end{equation}
From definition \eqref{eqn_Asym} it is clear that, a symmetric shock wave profile will have a density asymmetry quotient of unity, while for realistic shock waves its value is around unity and asymmetric shock profiles are predicted in experiments for strong hypersonic shock waves for which $Q_{\rho}$ is always grater than unity.

\begin{figure}
\includegraphics[width=8.5cm]{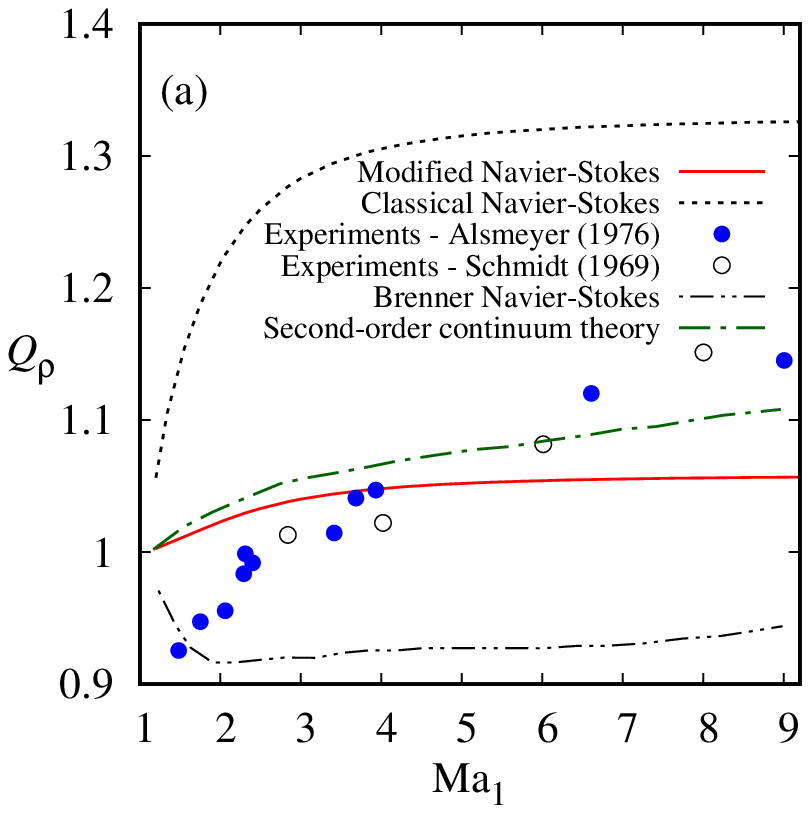}\,
\includegraphics[width=8.5cm]{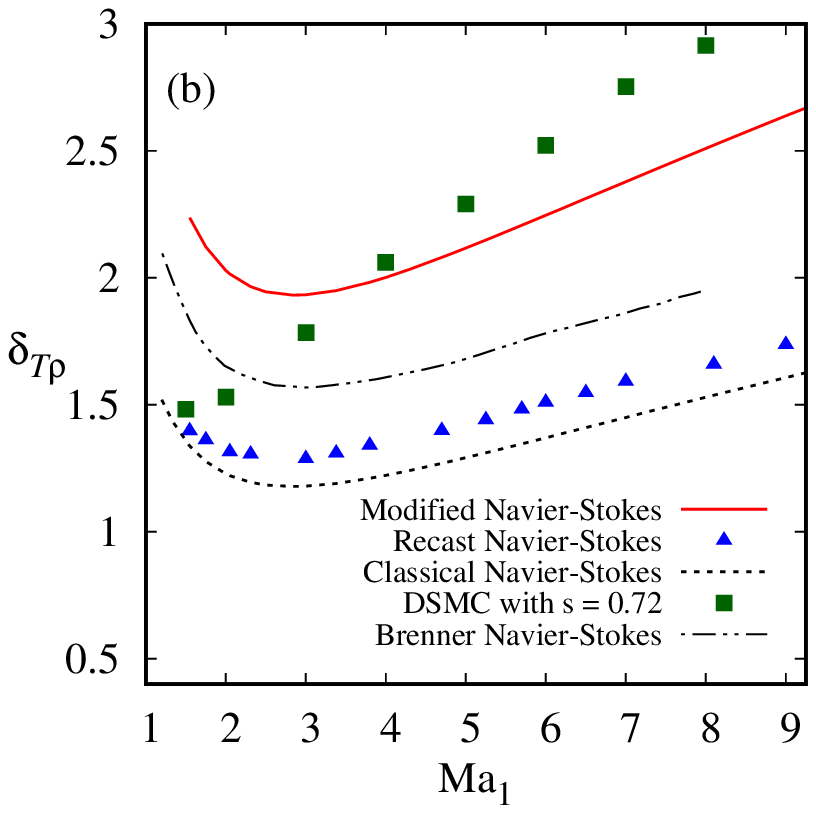}
\caption{(a) Variation of density asymmetry or shape factor, $Q_\rho$, as a function of upstream Mach number, $\rm{Ma}_1$, for argon gas. (b) Variation of temperature-density spatial lag, $\delta_{T\rho}$, as a function of upstream Mach number, $\rm{Ma}_1$, for argon gas. Details on the description of plotted data is indicated in Table~\ref{table:data}.}
\label{fig:7}
\end{figure}

Predictions of the modified Navier-Stokes equations for the density asymmetry quotient $Q_{\rho}$ are compared with experimental data of \citet{Alsmeyer1976} and \citet{Schmidt1969} along with the other theoretical predictions from the classical Navier-Stokes, a second-order continuum theory of \citet{Paolucci2018} and Brenner-Navier-Stokes equations in Fig.~\ref{fig:7}(a). One can observe that the classical Navier-Stokes equations predict an asymmetry quotient of more than unity (which means that the upstream part of the shock profile is more skewed than downstream) at all upstream Mach numbers and are far away from the experimental predictions. Brenner Navier-Stokes equations predicts $Q_{\rho} < 1$ (meaning that the downstream part of the shock profile is more skewed than the upstream part) for all upstream Mach numbers studied, however, predicted values for $Q_{\rho}$ show a close agreement with the experimental values for only weak shocks ($\rm{Ma}_1 \sim 1$). In contrast to the predictions of classical and Brenner Navier-Stokes, it is experimentally reported that the density profile has a significant asymmetry ($Q_{\rho} = 1 \pm 0.15$) at all upstream Mach numbers. The modified Navier-Stokes predicts an asymmetry quotient of around unity at low upstream Mach numbers and its value increases with shock strength. It is found that $Q_{\rho}$ values predicted by the modified NS equations fall within $8\%$ deviation from the experimental data and within $5\%$ deviation from the prediction by the second-order continuum theory \citet{Paolucci2018} which is evident from panel (a) of Fig.~\ref{fig:7}. It can be seen that the current modified Navier-Stokes and the second-order continuum theory of \citet{Paolucci2018} are better in predicting reasonable density asymmetry factor at all upstream Mach number.

Apart from the widely used shock macroscopic parameters, inverse shock thickness and density asymmetry factor, there is another important shock structure parameter which is defined based on the spatial difference between the temperature and density shock profiles, called density-temperature separation, $\delta_{T\rho}$. It is well-known that the variation in density and temperature within a shock do not occur at the same time due to the different finite relaxation times between momentum and energy transports. Spatial temperature variations occur well before spatial variations of density profile. The spatial difference between the normalized density and temperature profiles is denoted by $\delta_{T\rho}$ and is defined as,
\begin{equation}
\label{eqn_Trhosep}
\delta_{T\rho} = \arrowvert x(0.5\, T_{\rm N}) - x(0.5\, \rho_{\rm N}) \arrowvert.
\end{equation}
From definition \eqref{eqn_Trhosep} it is clear that the temperature-density separation measures the distance between the midpoints of normalized temperature and density profiles. Predictions by the different hydrodynamic theories for $\delta_{T\rho}$ are usually compared with available DSMC data \cite{LC1992,Greenshields2007}, due to lack of experimental data. Figure \ref{fig:7}(b) shows the comparison of shock macroscopic parameter temperature-density separation, $\delta_{T\rho}$, obtained from modified Navier-Stokes, Brenner-Navier-Stokes, recast Navier-Stokes and classical Navier-Stokes equations with DSMC data of \citet{LC1992}. It can be seen from panel (b) of Fig.~\ref{fig:7} that the DSMC data with a viscosity-temperature exponent $s = 0.72$ shows that $\delta_{T\rho}$ value increases with increasing Mach number, in particular increases from $\approx 1.5 \lambda_1$ to $\approx 2.9 \lambda_1$ when Mach number increases from $1.5$ to $8$. One can observe that predictions from the classical Navier-Stokes and recast Navier-Stokes equations increasingly under-predict $\delta_{T\rho}$ with increasing shock strength, while Brenner Navier-Stokes equations over-predict $\delta_{T\rho}$ for upstream Mach numbers, $\rm{Ma}_1 < 2$ and then increasingly under-predict from there. All hydrodynamic models presented here show a decreasing trend of $\delta_{T\rho}$ for $1.5\leq \rm{Ma}_1 \leq 3$ and then shows increasing trend of $\delta_{T\rho}$ from around $\rm{Ma}_1 > 3$. Modified Navier-Stokes equations shows the closest agreement with the DSMC data especially at high upstream Mach numbers as seen in panel (b) of Fig.~\ref{fig:7}. 

\section{\label{sec:V} Entropy generation within the shock layer}
Here we present and analyse the overall entropy generation and entropy profiles within the shock layer. From classical fluid theory, we know that the specific entropy $s$ is an increasing function of internal energy $e_{in}$ and these two thermodynamic quantities are interrelated via the Gibbs equation:
\begin{equation}
\label{eqn_Gibbs}
 \rho \, T \frac{D s}{D t} = \rho \frac{D e_{in}}{D t} + p\,\rho \frac{D }{D t} \left(\frac{1}{\rho}\right),
\end{equation}
where $D/D t = \partial /\partial t + U \bcdot \nabla $ denotes a material derivative.
As a consequence of the Gibbs equation \eqref{eqn_Gibbs}, the energy balance equation \eqref{eqn_energy} can be replaced by an equation on the entropy $s = c_v\, \ln  (T/\rho^{\gamma-1})$
and is obtained as:
\begin{equation}
\label{eqn_Entropy}
\rho \frac{D s}{D t} + \nabla \bcdot \left( \frac{\mbfq}{T} \right) = \dot{s}_{gen},
\end{equation}
where $\dot{s}_{gen}$ denotes the rate of entropy generation and is given as:
\begin{equation}
 \dot{s}_{gen} = \frac{1}{T} \left[ \mbfpi \colon \nabla U \,-\, \frac{1}{T} \left( \mbfq \bcdot \nabla T \right)\right].
\end{equation}
It is note worthy to mention here that equation \eqref{eqn_Entropy} serves as a mathematical formulation of the second law of thermodynamics and accordingly the entropy generation rate $\dot{s}_{gen}$ should be a non-negative quantity for any physically admissible process. The nondimensionalized entropy generation rate inside the shock layer for the modified Navier-Stokes equations can be obtained as:

\begin{figure}[ht]
\includegraphics[width=7.5cm]{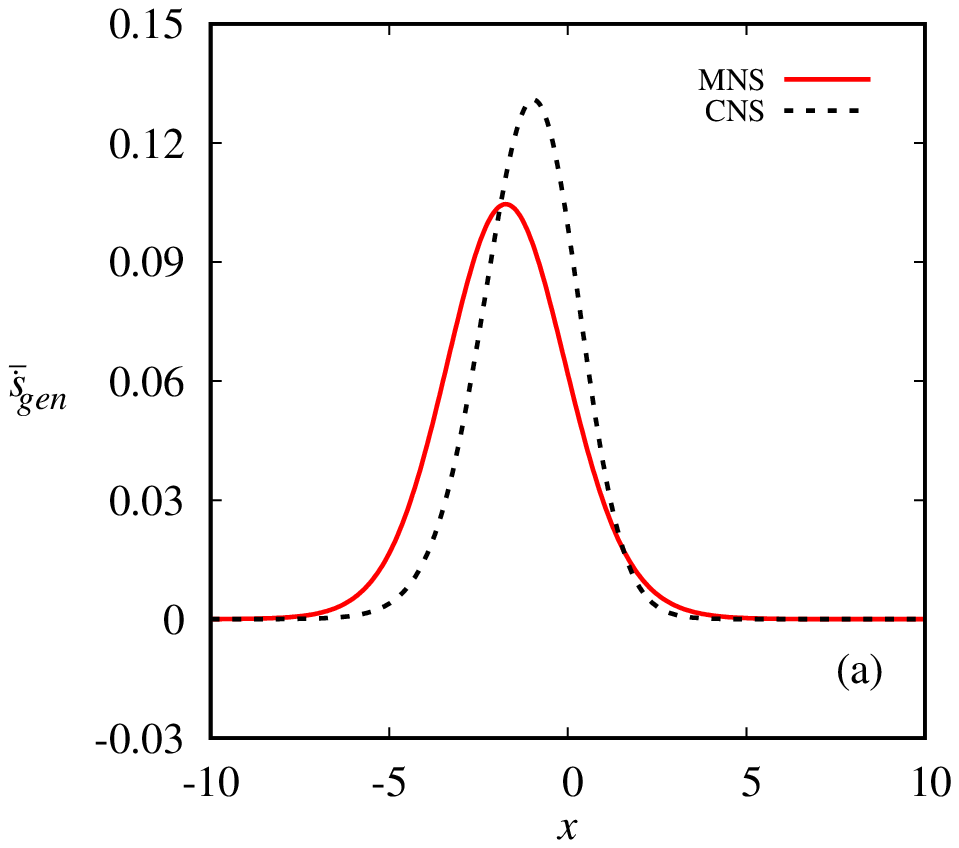}\,
\includegraphics[width=7.5cm]{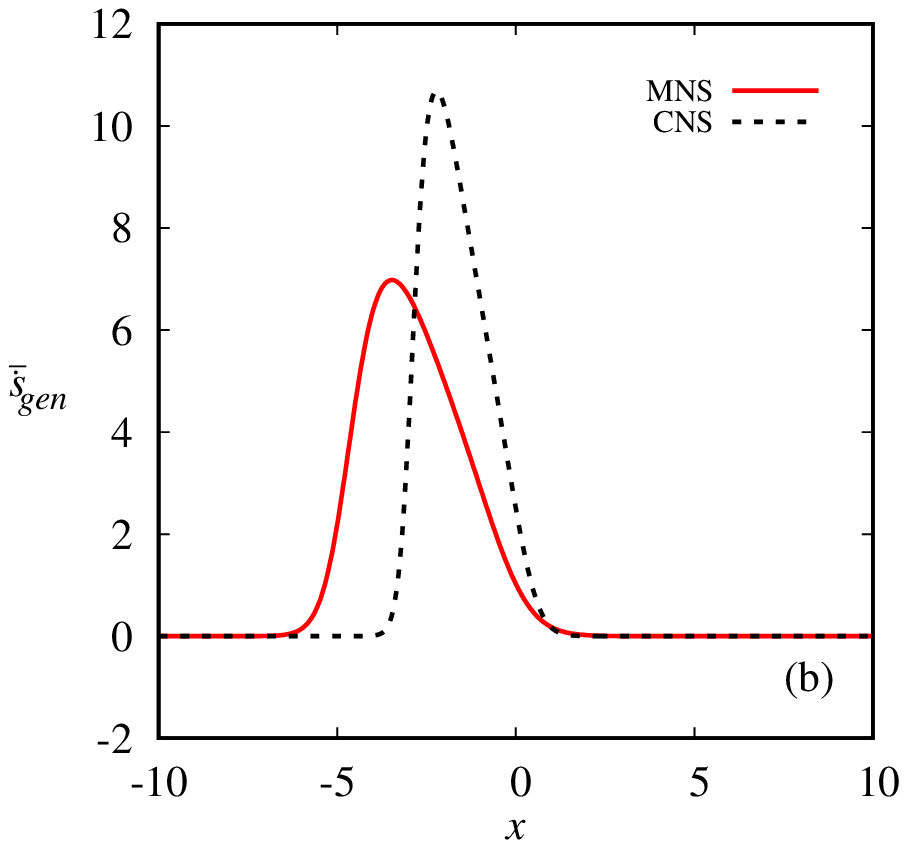}
\includegraphics[width=7.5cm]{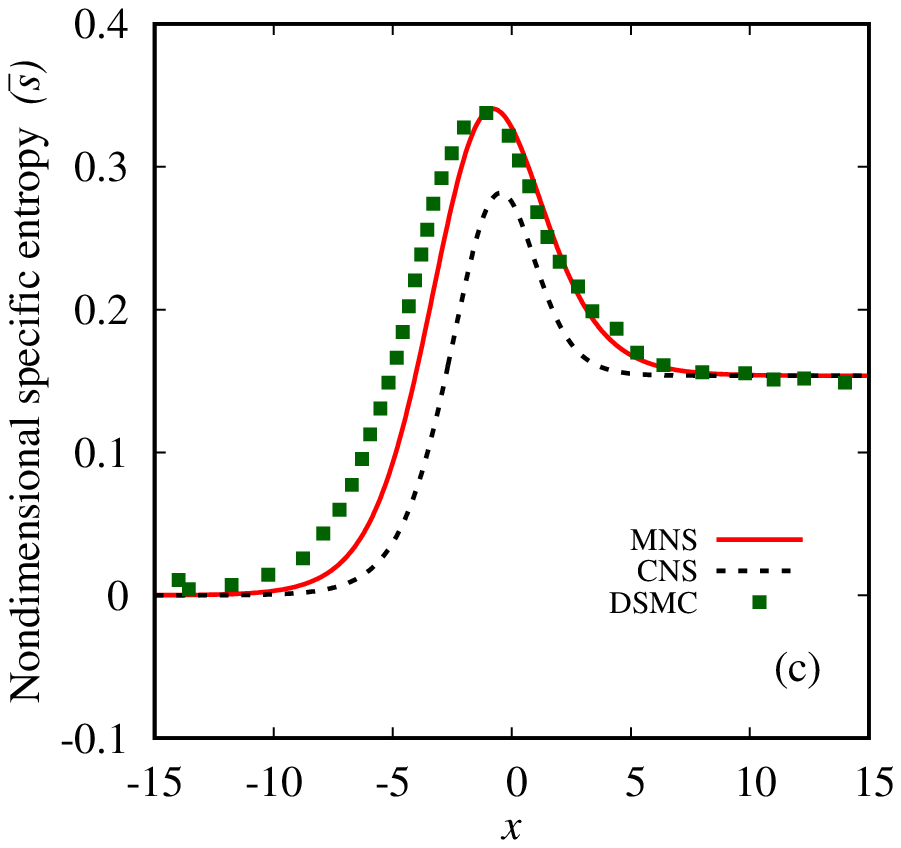}\,
\includegraphics[width=7.5cm]{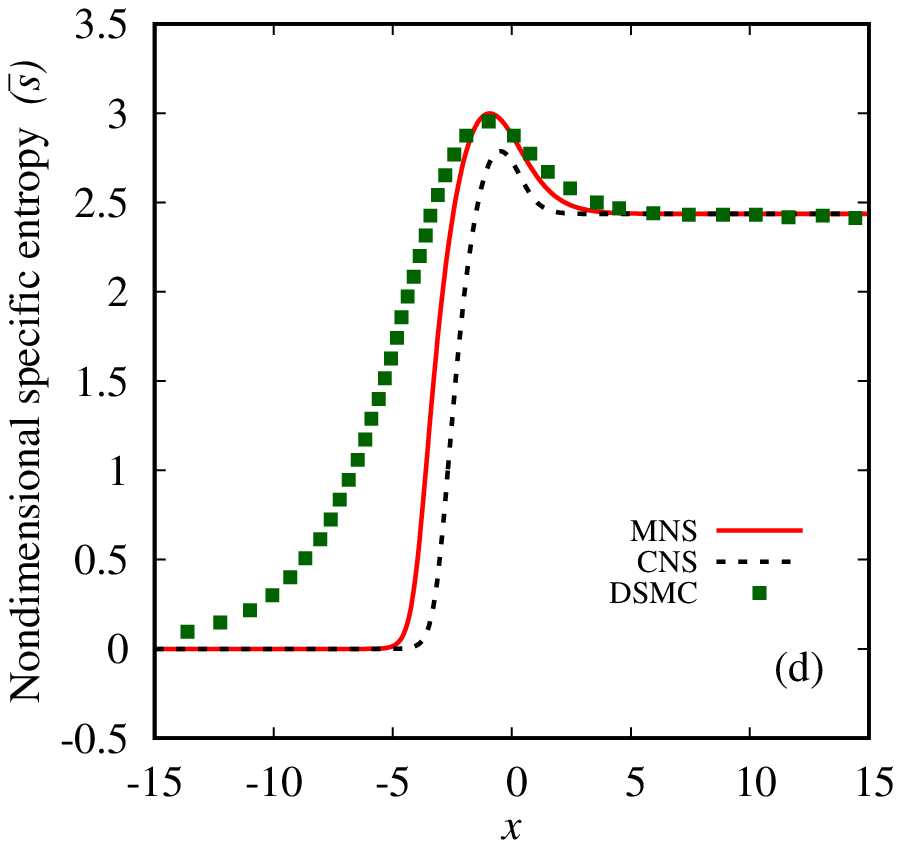}
\caption{(a,b) Nondimensional entropy production rate $(\overline{\dot{s}}_{gen})$ and (c,d) nondimensional specific entropy profiles within the shock: for (a,c) $\rm{Ma}_1 = 1.75$ and (b,d) $\rm{Ma}_1 = 6$. In each panel, the dashed line (black) and the solid line (red) refer to solutions from classical Navier-Stokes (CNS) and modified Navier-Stokes equations (MNS), respectively. The available DSMC data from \citet{Schrock2005} are superimposed for the specific entropy profiles in panels (c) and (d) as filled squares (green).}
\label{fig:8}
\end{figure}

\begin{eqnarray}
\label{eqn_Entropygen}
\overline{\dot{s}}_{gen} &=& \underline{\frac{4}{3} \frac{\overline{\mu}}{\overline{T}} \left( \frac{{\rm d} \overline{u}}{{\rm d} \overline{x}} \right)^2} + \overline{ \frac{4}{3} \left( \frac{\gamma}{\la_0}\right) \frac{\overline{\mu}}{\overline{T}} \frac{\overline{\ka}_m}{\overline{\rho}} \frac{{\rm d}^2 \overline{\rho}}{{\rm d} \overline{x}^2} \frac{{\rm d} \overline{u}}{{\rm d} \overline{x}} } - \frac{4}{3} \left( \frac{\gamma}{\la_0}\right) \frac{\overline{\mu}}{\overline{T}} \frac{\overline{\ka}_m}{\overline{\rho^2}}  \left( \frac{{\rm d} \overline{\rho}}{{\rm d} \overline{x}} \right)^2  \frac{{\rm d} \overline{u}}{{\rm d} \overline{x}}  \nonumber \\
&& + \underline{\frac{\overline{\ka}}{\overline{T}^2} \left(\frac{{\rm d} \overline{T}}{{\rm d} \overline{x}}\right)^2}+ \frac{\gamma}{\left(\gamma - 1\right)\, \rm{Pr}} \frac{\overline{\ka}_m}{\overline{T}} \frac{{\rm d} \overline{\rho}}{{\rm d} \overline{x}} \frac{{\rm d} \overline{T}}{{\rm d} \overline{x}}.
\end{eqnarray}

Setting $\overline{\ka}_m = 0$ reduces Eq.~\eqref{eqn_Entropygen} to the underlined terms only and represents the rate of entropy production for the classical Navier-Stokes equations which appears always non-negative. To justify the non-negativity of the entropy production rate for the modified continuum equations one observes the following in Eq.~\eqref{eqn_Entropygen}. Across the shock layer, the gradient of velocity is always negative as the velocity progressively decreases from supersonic/hypersonic upstream to subsonic downstream. The gradients of temperature and density are always positive due to the fact that they progressively increase within the shock. Consequently, all terms represented in Eq.~\eqref{eqn_Entropygen} are always positive throughout the shock layer except the over lined term. This term representing the second-order derivative of density, $\left( {\rm d}^2 \overline{\rho} / {\rm d} \overline{x}^2 \right)$ in   Eq.~\eqref{eqn_Entropygen}, changes sign from positive to negative within the shock layer. However, contribution to the entropy production due to this term appears negligible when the full expression in Eq.~\eqref{eqn_Entropygen} is evaluated within the shock layer. This results in the non-negativity of the full entropy generation rate, $\overline{\dot{s}}_{gen}$, at all Mach number. This is illustrated in Fig.~\ref{fig:8} (a) and (b) for two different upstream Mach numbers $\rm{Ma}_1 = 1.75$ and $\rm{Ma}_1 = 6$, respectively. Moreover, on these figures one observes that the modified Navier-Stokes generates entropy in a wider region compared to the classical. This means that the modified Navier-Stokes captures more non-equilibrium effects in the shock region than the classical \cite{Schrock2005}. Figure~\ref{fig:8} (c) and (d) then shows the nondimensional specific entropy $\overline{s} = (1/(\gamma-1)) \ln \left(\overline{T}/\overline{\rho}^{\gamma-1}\right)$ profiles across the shock layer for upstream Mach numbers of $\rm{Ma}_1 = 1.75$ and $\rm{Ma}_1 = 6$. DSMC data of the specific entropy at the same Mach numbers from \citet{Schrock2005} are also plotted on these figures. From panels (c) and (d) of Fig.~\ref{fig:8} one can observe that the entropy maximum and / or peak is occurring within the shock layer. While all models approximately predict the same spatial location for the peak, the magnitude of the peak in the modified Navier-Stokes is higher than in the classical. The modified Navier-Stokes matches the DSMC prediction of the magnitude of the peak. Overall the DSMC high specific entropy profiles corroborate the modified Navier-Stokes with its  wider entropy generation region as seen upstream of the shock on panels (a) and (b) of of Fig.~\ref{fig:8} \cite{Schrock2005}.
\section{\label{sec:VI}Conclusions}

In this work we presented a full numerical investigation into shock wave profile description in a monatomic gas using hydrodynamics models by identifying constitutive equations that provide better agreement for all parameters involved in the prediction of shock structures. We have shown that our identified constitutive equations allow for non-negative entropy production rate across the shock for the modified hydrodynamic equations. A detailed comparison between the predictions of the modified hydrodynamic equations with Alsmeyer's experimental data and DSMC data along with classical Navier-Stokes hydrodynamic solutions is presented for upstream Mach number ranging between $1.5$ and $11$. First, we focused on the shock density profiles as accurate data from the experiments are available. We then included comparison of shock temperature profiles with available DSMC data. Finally, we showed the comparison of all three well-known shock macroscopic parameters (inverse shock thickness, density asymmetry factor and temperature-density separation) with available experimental and DSMC data. Our analysis showed that the modified hydrodynamic equations provide excellent quantitative agreement with experimental data as well as with DSMC data at all Mach numbers discussed and especially best in reproducing experimental trends for the shock density and inverse shock thickness profiles. In fact we conclude that the results are improvement upon those obtained previously in bi-velocity hydrodynamics and more accurate than those obtained using equations from the extended hydrodynamic approach of kinetic theory. Further implications of these results as related to recently proposed recast Navier-Stokes equations are still to be investigated. However, the presently identified constitutive equations that are of volume/mass diffusion type may be specifically adopted in a multi-scale simulation for better shock structure descriptions. 

{\bf Acknowledgement:}
This research is funded in the UK by the Engineering and Physical Sciences Research Council (EPSRC) under grant no. EP/R008027/1 and the Leverhulme Trust under grant Ref. RPG-2018-174.
\appendix
\section{Dimensional analysis of the modified continuum flow model}
\label{AppendixA}
The one-dimensional modified Navier-Stokes model can be written in `conservative' form as:
\begin{align}
& \frac{\partial \rho}{\partial t} + \frac{\partial}{\partial x} \Big(\, \rho \, u \Big) = 0, \label{eqn_D1}\\
& \frac{\partial}{\partial t} \Big( \rho \, u \Big) + \frac{\partial}{\partial x} \Big(\, \rho \, u^2 + \, \rho \, \bR \, T\,  +\, \Pi\, \Big)  = 0,\label{eqn_D2}\\
& \frac{\partial}{\partial t} \left(\frac{1}{2}\rho\,  u^2\,  + \, \frac{1}{(\gamma-1)} \, \rho \, \bR \, T \right) + \frac{\partial}{\partial x} \left(\frac{1}{2}  \rho \, u^3 \, + \frac{\gamma}{(\gamma-1)} \, \rho \, \bR \,T \,u \, + \, \Pi \, u \, +\,  q \, \right) = 0, \label{eqn_D3}
\end{align}
with constitutive relations for the longitudinal stress $\Pi$ and the heat flux $q$ as:  
\begin{align}
& \Pi = -\,\frac{4}{3}\, \mu \, \frac{\partial u}{\partial x} \,-\,\frac{4}{3} \frac{\mu\,\kam}{\rho} \, \frac{\partial^2 \rho}{\partial x^2}\,+\,\frac{4}{3} \frac{\mu\,\kam}{\rho^2} \left( \frac{\partial \rho}{\partial x} \right)^2, \label{eqn_D4}\\
& q = - \ka \, \frac{\partial T}{\partial x} - \frac{\gamma}{(\gamma-1) \rm{Pr}} \kam \,\bR \, T \,\frac{\partial \rho}{\partial x}. \label{eqn_D5}
\end{align}

For a shock wave structure in a monatomic gas, let us assume the characteristic length scale to be $L$, the speed of sound being the natural choice for the velocity scale $U$ ($\sim \sqrt{\gamma\,\bR \, T_{ref}}$) and $L/U$ to be the natural advection time scale. With these scales, we  introduce the following dimensionless variables:
\begin{align}
\begin{split}
\label{eqn_D7}
& \overline{x} = \frac{x}{L}, \quad \overline{u} = \frac{u}{U},\quad \overline{t} = \frac{t U}{L}, \quad \overline{T} = \frac{\bR}{U^2} T,\quad \overline{\rho} = \frac{\rho}{\rho_{ref}} , \quad \overline{\mu} = \frac{\mu}{\mu_{ref}}.
\end{split}
\end{align}
Using the scaled variables given in Eq.~\ref{eqn_D7}, the one-dimensional conservation equations become:
\begin{align}
 & \frac{\partial \overline{\rho}}{\partial \overline{t}} + \frac{\partial}{\partial \overline{x}} \Big(\, \overline{\rho} \, \overline{u} \Big) = 0, \label{eqn_D8}  \\
 & \frac{\partial}{\partial \overline{t}} \Big( \overline{\rho} \, \overline{u} \Big) +  \frac{\partial}{\partial \overline{x}} \Big(\, \overline{\rho} \, \overline{u}^2 \,+\,\overline{\rho} \, \overline{T} \Big)  =  \frac{\mu_{ref}}{\rho_{ref}\, U\,L} \frac{\partial}{\partial \overline{x}} \left( \frac{4}{3} \overline{\mu} \frac{\partial \overline{u}}{\partial \overline{x}} \right) \nonumber \\
 & \qquad \qquad \qquad \qquad \qquad \qquad +  \left(\frac{\mu_{ref}}{\rho_{ref}\, U\,L} \right)^2 \frac{\partial}{\partial \overline{x}}  \left( \frac{4}{3} \kappa_{m_0} \frac{\overline{\mu}^2}{\overline{\rho}^2} \frac{\partial^2 \overline{\rho}}{\partial \overline{x}^2}\, - \,\frac{4}{3} \kappa_{m_0} \frac{\overline{\mu}^2}{\overline{\rho}^3} \left( \frac{\partial \overline{\rho}}{\partial \overline{x}} \right)^2 \right),  \label{eqn_D9}\\
 &  \frac{\partial}{\partial \overline{t}} \left(\frac{1}{2} \overline{\rho}\,  \overline{u}^2\, + \, \frac{1}{(\gamma-1)} \,\overline{\rho} \, \overline{T} \right) \, + \frac{\partial}{\partial \overline{x}} \left(\frac{1}{2}  \overline{\rho} \, \overline{u}^3 \, +\, \frac{\gamma}{(\gamma-1)} \, \overline{\rho} \, \overline{T} \,\overline{u} \right)  \nonumber \\
 &  \qquad  = \frac{\mu_{ref}}{\rho_{ref}\, U\,L} \frac{\partial}{\partial \overline{x}} \left( \frac{4}{3} \overline{\mu} \, \overline{u} \frac{\partial \overline{u}}{\partial \overline{x}} \right) \,+\,\left(\frac{\mu_{ref}}{\rho_{ref}\, U\,L} \right)^2 \frac{\partial}{\partial \overline{x}}  \left( \frac{4}{3} \kappa_{m_0} \frac{\overline{\mu}^2}{\overline{\rho}^2} \overline{u} \frac{\partial^2 \overline{\rho}}{\partial \overline{x}^2}\, - \,\frac{4}{3} \kappa_{m_0} \frac{\overline{\mu}^2}{\overline{\rho}^3} \overline{u} \left( \frac{\partial \overline{\rho}}{\partial \overline{x}} \right)^2 \right) \nonumber \\ 
 & \qquad \qquad + \frac{\mu_{ref}}{\rho_{ref}\, U\,L} \frac{\partial}{\partial \overline{x}} \left( \overline{\ka} \frac{\partial \overline{T}}{\partial \overline{x}}\,+\, \kappa_{m_0} \overline{\ka}  \frac{\overline{T}}{\overline{\rho}}\frac{\partial \overline{\rho}}{\partial \overline{x}}  \right).  \label{eqn_D10}
\end{align}

It is well-known that shock structure is an example of highly non-equilibrium and rarefied gas flow, hence, the generally accepted dimensionless parameter which measures the gas rarefaction is the Knudsen number ($\rm Kn$) and is defined as the ratio of the mean free path ($\la$) of the gas molecules to the characteristic length scale ($L$) of the system: ${\rm Kn} = \la/L$.
The mean free path of the gas is proportional to the viscosity coefficient and inversely proportional to the density and the square root of the temperature of the gas molecules. Using the current scales the mean free path of gas is given as:
\begin{equation}
 \label{eqn_D12}
 \la  \propto \frac{\mu_{ref}}{\rho_{ref}\, \sqrt{\gamma\,\bR\, T_{ref}}} = \frac{\mu_{ref}}{\rho_{ref}\, U}.
\end{equation}
Finally, the dimensionless conservation equations involving Knudsen number as dimensional parameter are:
\begin{align}
 & \frac{\partial \overline{\rho}}{\partial \overline{t}} + \frac{\partial}{\partial \overline{x}} \Big(\, \overline{\rho} \, \overline{u} \Big) = 0, \label{eqn_D13}\\  
 & \frac{\partial}{\partial \overline{t}} \Big( \overline{\rho} \, \overline{u} \Big) +  \frac{\partial}{\partial \overline{x}} \Big(\, \overline{\rho} \, \overline{u}^2 \,+\,\overline{\rho} \, \overline{T} \Big)  =  {\rm Kn} \frac{\partial}{\partial \overline{x}} \left( \frac{4}{3} \overline{\mu} \frac{\partial \overline{u}}{\partial \overline{x}} \right) \nonumber \\ 
 & \qquad \qquad \qquad \qquad \qquad \qquad \quad +  {\rm Kn}^2 \frac{\partial}{\partial \overline{x}}  \left( \frac{4}{3} \kappa_{m_0} \frac{\overline{\mu}^2}{\overline{\rho}^2} \frac{\partial^2 \overline{\rho}}{\partial \overline{x}^2}\, - \,\frac{4}{3} \kappa_{m_0} \frac{\overline{\mu}^2}{\overline{\rho}^3} \left( \frac{\partial \overline{\rho}}{\partial \overline{x}} \right)^2 \right),  \label{eqn_D14}  \\
 &  \frac{\partial}{\partial \overline{t}} \left(\frac{1}{2} \overline{\rho}\,  \overline{u}^2\, + \, \frac{1}{(\gamma-1)} \,\overline{\rho} \, \overline{T} \right) \, + \frac{\partial}{\partial \overline{x}} \left(\frac{1}{2}  \overline{\rho} \, \overline{u}^3 \, +\, \frac{\gamma}{(\gamma-1)} \, \overline{\rho} \, \overline{T} \,\overline{u} \right) \nonumber \\
 & \qquad \qquad = {\rm Kn} \frac{\partial}{\partial \overline{x}} \left( \frac{4}{3} \overline{\mu} \, \overline{u} \frac{\partial \overline{u}}{\partial \overline{x}} \right) \,+\,{\rm Kn}^2 \frac{\partial}{\partial \overline{x}}  \left( \frac{4}{3} \kappa_{m_0} \frac{\overline{\mu}^2}{\overline{\rho}^2} \overline{u} \frac{\partial^2 \overline{\rho}}{\partial \overline{x}^2}\, - \,\frac{4}{3} \kappa_{m_0} \frac{\overline{\mu}^2}{\overline{\rho}^3} \overline{u} \left( \frac{\partial \overline{\rho}}{\partial \overline{x}} \right)^2 \right)  \nonumber \\
 & \qquad \qquad \quad + {\rm Kn} \frac{\partial}{\partial \overline{x}} \left( \overline{\ka} \frac{\partial \overline{T}}{\partial \overline{x}}\,+\, \kappa_{m_0} \overline{\ka}  \frac{\overline{T}}{\overline{\rho}}\frac{\partial \overline{\rho}}{\partial \overline{x}}  \right). \label{eqn_D15}
\end{align}

One observes that most new diffusion terms in the dimensionless conservation equations (\eqref{eqn_D13} - \eqref{eqn_D15}) may be neglected in the vanishing limit of Knudsen number (${\rm Kn} \rightarrow 0 $). However as the Knudsen number increases (as in the case of shock profile description) these terms may no longer be neglected.

\vspace*{0.5cm}


\begin{thebibliography}{0}%
\makeatletter
\providecommand \@ifxundefined [1]{%
 \@ifx{#1\undefined}
}%
\providecommand \@ifnum [1]{%
 \ifnum #1\expandafter \@firstoftwo
 \else \expandafter \@secondoftwo
 \fi
}%
\providecommand \@ifx [1]{%
 \ifx #1\expandafter \@firstoftwo
 \else \expandafter \@secondoftwo
 \fi
}%
\providecommand \natexlab [1]{#1}%
\providecommand \enquote  [1]{``#1''}%
\providecommand \bibnamefont  [1]{#1}%
\providecommand \bibfnamefont [1]{#1}%
\providecommand \citenamefont [1]{#1}%
\providecommand \href@noop [0]{\@secondoftwo}%
\providecommand \href [0]{\begingroup \@sanitize@url \@href}%
\providecommand \@href[1]{\@@startlink{#1}\@@href}%
\providecommand \@@href[1]{\endgroup#1\@@endlink}%
\providecommand \@sanitize@url [0]{\catcode `\\12\catcode `\$12\catcode
  `\&12\catcode `\#12\catcode `\^12\catcode `\_12\catcode `\%12\relax}%
\providecommand \@@startlink[1]{}%
\providecommand \@@endlink[0]{}%
\providecommand \url  [0]{\begingroup\@sanitize@url \@url }%
\providecommand \@url [1]{\endgroup\@href {#1}{\urlprefix }}%
\providecommand \urlprefix  [0]{URL }%
\providecommand \Eprint [0]{\href }%
\providecommand \doibase [0]{https://doi.org/}%
\providecommand \selectlanguage [0]{\@gobble}%
\providecommand \bibinfo  [0]{\@secondoftwo}%
\providecommand \bibfield  [0]{\@secondoftwo}%
\providecommand \translation [1]{[#1]}%
\providecommand \BibitemOpen [0]{}%
\providecommand \bibitemStop [0]{}%
\providecommand \bibitemNoStop [0]{.\EOS\space}%
\providecommand \EOS [0]{\spacefactor3000\relax}%
\providecommand \BibitemShut  [1]{\csname bibitem#1\endcsname}%
\let\auto@bib@innerbib\@empty
\end{thebibliography}%


\begin{thebibliography}{61}%
\makeatletter
\providecommand \@ifxundefined [1]{%
 \@ifx{#1\undefined}
}%
\providecommand \@ifnum [1]{%
 \ifnum #1\expandafter \@firstoftwo
 \else \expandafter \@secondoftwo
 \fi
}%
\providecommand \@ifx [1]{%
 \ifx #1\expandafter \@firstoftwo
 \else \expandafter \@secondoftwo
 \fi
}%
\providecommand \natexlab [1]{#1}%
\providecommand \enquote  [1]{``#1''}%
\providecommand \bibnamefont  [1]{#1}%
\providecommand \bibfnamefont [1]{#1}%
\providecommand \citenamefont [1]{#1}%
\providecommand \href@noop [0]{\@secondoftwo}%
\providecommand \href [0]{\begingroup \@sanitize@url \@href}%
\providecommand \@href[1]{\@@startlink{#1}\@@href}%
\providecommand \@@href[1]{\endgroup#1\@@endlink}%
\providecommand \@sanitize@url [0]{\catcode `\\12\catcode `\$12\catcode
  `\&12\catcode `\#12\catcode `\^12\catcode `\_12\catcode `\%12\relax}%
\providecommand \@@startlink[1]{}%
\providecommand \@@endlink[0]{}%
\providecommand \url  [0]{\begingroup\@sanitize@url \@url }%
\providecommand \@url [1]{\endgroup\@href {#1}{\urlprefix }}%
\providecommand \urlprefix  [0]{URL }%
\providecommand \Eprint [0]{\href }%
\providecommand \doibase [0]{https://doi.org/}%
\providecommand \selectlanguage [0]{\@gobble}%
\providecommand \bibinfo  [0]{\@secondoftwo}%
\providecommand \bibfield  [0]{\@secondoftwo}%
\providecommand \translation [1]{[#1]}%
\providecommand \BibitemOpen [0]{}%
\providecommand \bibitemStop [0]{}%
\providecommand \bibitemNoStop [0]{.\EOS\space}%
\providecommand \EOS [0]{\spacefactor3000\relax}%
\providecommand \BibitemShut  [1]{\csname bibitem#1\endcsname}%
\let\auto@bib@innerbib\@empty
\bibitem [{\citenamefont {Courant}\ and\ \citenamefont
  {Friedrichs}(1948)}]{CF1948}%
  \BibitemOpen
  \bibfield  {author} {\bibinfo {author} {\bibfnamefont {R.}~\bibnamefont
  {Courant}}\ and\ \bibinfo {author} {\bibfnamefont {K.~O.}\ \bibnamefont
  {Friedrichs}},\ }\href@noop {} {\emph {\bibinfo {title} {Supersonic Flow and
  Shock Waves}}}\ (\bibinfo  {publisher} {New York: Interscience},\ \bibinfo
  {year} {1948})\BibitemShut {NoStop}%
\bibitem [{\citenamefont {Liepmann}\ and\ \citenamefont
  {Roshko}(1957)}]{LR1957}%
  \BibitemOpen
  \bibfield  {author} {\bibinfo {author} {\bibfnamefont {H.~W.}\ \bibnamefont
  {Liepmann}}\ and\ \bibinfo {author} {\bibfnamefont {A.}~\bibnamefont
  {Roshko}},\ }\href@noop {} {\emph {\bibinfo {title} {Elements of {G}as
  {D}ynamics}}}\ (\bibinfo  {publisher} {John Wiley \& Sons, Inc., New York},\
  \bibinfo {year} {1957})\BibitemShut {NoStop}%
\bibitem [{\citenamefont {Grad}(1952)}]{Grad1952}%
  \BibitemOpen
  \bibfield  {author} {\bibinfo {author} {\bibfnamefont {H.}~\bibnamefont
  {Grad}},\ }\bibfield  {title} {\bibinfo {title} {The profile of a steady
  plane shock wave},\ }\href {https://doi.org/10.1002/cpa.3160050304}
  {\bibfield  {journal} {\bibinfo  {journal} {Commun.~Pure Appl.~Math.}\
  }\textbf {\bibinfo {volume} {5}},\ \bibinfo {pages} {257} (\bibinfo {year}
  {1952})}\BibitemShut {NoStop}%
\bibitem [{\citenamefont {Bird}(1994)}]{Bird1994}%
  \BibitemOpen
  \bibfield  {author} {\bibinfo {author} {\bibfnamefont {G.~A.}\ \bibnamefont
  {Bird}},\ }\href@noop {} {\emph {\bibinfo {title} {Molecular Gas Dynamics and
  the Direct Simulation of Gas Flows}}}\ (\bibinfo  {publisher} {Oxford Univ
  Press},\ \bibinfo {year} {1994})\BibitemShut {NoStop}%
\bibitem [{\citenamefont {Reese}\ \emph {et~al.}(1995)\citenamefont {Reese},
  \citenamefont {Woods}, \citenamefont {Thivet},\ and\ \citenamefont
  {Candel}}]{Reeseetal1995}%
  \BibitemOpen
  \bibfield  {author} {\bibinfo {author} {\bibfnamefont {J.~M.}\ \bibnamefont
  {Reese}}, \bibinfo {author} {\bibfnamefont {L.~C.}\ \bibnamefont {Woods}},
  \bibinfo {author} {\bibfnamefont {F.~J.~P.}\ \bibnamefont {Thivet}},\ and\
  \bibinfo {author} {\bibfnamefont {S.~M.}\ \bibnamefont {Candel}},\ }\bibfield
   {title} {\bibinfo {title} {A second-order description of shock structure},\
  }\href {https://doi.org/https://doi.org/10.1006/jcph.1995.1062} {\bibfield
  {journal} {\bibinfo  {journal} {J.~Comput.~Phys.}\ }\textbf {\bibinfo
  {volume} {117}},\ \bibinfo {pages} {240} (\bibinfo {year}
  {1995})}\BibitemShut {NoStop}%
\bibitem [{\citenamefont {LeVeque}(2002)}]{LeVeque2002}%
  \BibitemOpen
  \bibfield  {author} {\bibinfo {author} {\bibfnamefont {R.~J.}\ \bibnamefont
  {LeVeque}},\ }\href@noop {} {\emph {\bibinfo {title} {Finite Volume Methods
  for Hyperbolic Problems}}}\ (\bibinfo  {publisher} {Cambridge University
  Press},\ \bibinfo {year} {2002})\BibitemShut {NoStop}%
\bibitem [{\citenamefont {Greenshields}\ and\ \citenamefont
  {Reese}(2007)}]{Greenshields2007}%
  \BibitemOpen
  \bibfield  {author} {\bibinfo {author} {\bibfnamefont {C.~J.}\ \bibnamefont
  {Greenshields}}\ and\ \bibinfo {author} {\bibfnamefont {J.~M.}\ \bibnamefont
  {Reese}},\ }\bibfield  {title} {\bibinfo {title} {The structure of shock
  waves as a test of {B}renner's modifications to the {N}avier-{S}tokes
  equations},\ }\href {https://doi.org/10.1017/S0022112007005575} {\bibfield
  {journal} {\bibinfo  {journal} {J.~Fluid Mech.}\ }\textbf {\bibinfo {volume}
  {580}},\ \bibinfo {pages} {407} (\bibinfo {year} {2007})}\BibitemShut
  {NoStop}%
\bibitem [{\citenamefont {Reddy}\ and\ \citenamefont {Alam}(2015)}]{RA2015}%
  \BibitemOpen
  \bibfield  {author} {\bibinfo {author} {\bibfnamefont {M.~H.~L.}\
  \bibnamefont {Reddy}}\ and\ \bibinfo {author} {\bibfnamefont
  {M.}~\bibnamefont {Alam}},\ }\bibfield  {title} {\bibinfo {title} {Plane
  shock waves and {H}aff's law in a granular gas},\ }\href
  {https://doi.org/10.1017/jfm.2015.455} {\bibfield  {journal} {\bibinfo
  {journal} {J.~Fluid Mech.}\ }\textbf {\bibinfo {volume} {779}},\ \bibinfo
  {pages} {R2} (\bibinfo {year} {2015})}\BibitemShut {NoStop}%
\bibitem [{\citenamefont {Reddy}(2016)}]{Reddy2016}%
  \BibitemOpen
  \bibfield  {author} {\bibinfo {author} {\bibfnamefont {M.~H.~L.}\
  \bibnamefont {Reddy}},\ }\emph {\bibinfo {title} {Plane shock waves in
  granular gases and regularized moment equations}},\ \href@noop {} {Ph.D.
  thesis},\ \bibinfo  {school} {Jawaharlal Nehru Centre for Advanced Scientific
  Research, India} (\bibinfo {year} {2016})\BibitemShut {NoStop}%
\bibitem [{\citenamefont {Reddy}\ and\ \citenamefont
  {Alam}(2016{\natexlab{a}})}]{RA2016}%
  \BibitemOpen
  \bibfield  {author} {\bibinfo {author} {\bibfnamefont {M.~H.~L.}\
  \bibnamefont {Reddy}}\ and\ \bibinfo {author} {\bibfnamefont
  {M.}~\bibnamefont {Alam}},\ }\bibfield  {title} {\bibinfo {title} {Plane
  shock wave structure in a dilute granular gas},\ }\href
  {https://doi.org/10.1063/1.4967625} {\bibfield  {journal} {\bibinfo
  {journal} {AIP Conf. Proc.}\ }\textbf {\bibinfo {volume} {1786}},\ \bibinfo
  {pages} {120001} (\bibinfo {year} {2016}{\natexlab{a}})}\BibitemShut
  {NoStop}%
\bibitem [{\citenamefont {Von~Mises}(1950)}]{Mises1950}%
  \BibitemOpen
  \bibfield  {author} {\bibinfo {author} {\bibfnamefont {R.}~\bibnamefont
  {Von~Mises}},\ }\bibfield  {title} {\bibinfo {title} {On the thickness of a
  steady shock wave},\ }\href {https://doi.org/10.2514/8.1723} {\bibfield
  {journal} {\bibinfo  {journal} {J. Aeronaut. Sci.}\ }\textbf {\bibinfo
  {volume} {17}},\ \bibinfo {pages} {551} (\bibinfo {year} {1950})}\BibitemShut
  {NoStop}%
\bibitem [{\citenamefont {Gilbarg}\ and\ \citenamefont
  {Paolucci}(1953)}]{GP1953}%
  \BibitemOpen
  \bibfield  {author} {\bibinfo {author} {\bibfnamefont {D.}~\bibnamefont
  {Gilbarg}}\ and\ \bibinfo {author} {\bibfnamefont {D.}~\bibnamefont
  {Paolucci}},\ }\bibfield  {title} {\bibinfo {title} {The structure of shock
  waves in the continuum theory of fluids},\ }\href@noop {} {\bibfield
  {journal} {\bibinfo  {journal} {J.~Rat.~Mech.~Anal.}\ }\textbf {\bibinfo
  {volume} {2}},\ \bibinfo {pages} {617} (\bibinfo {year} {1953})}\BibitemShut
  {NoStop}%
\bibitem [{\citenamefont {Narasimha}\ and\ \citenamefont
  {Das}(1990)}]{NarasimhaDas1990}%
  \BibitemOpen
  \bibfield  {author} {\bibinfo {author} {\bibfnamefont {R.}~\bibnamefont
  {Narasimha}}\ and\ \bibinfo {author} {\bibfnamefont {P.}~\bibnamefont
  {Das}},\ }\bibfield  {title} {\bibinfo {title} {A spectral solution of the
  {B}oltzmann equation for the infinitely strong shock},\ }\href@noop {}
  {\bibfield  {journal} {\bibinfo  {journal} {Philos. Trans. R. Soc. London,
  Ser. A}\ }\textbf {\bibinfo {volume} {330}},\ \bibinfo {pages} {217}
  (\bibinfo {year} {1990})}\BibitemShut {NoStop}%
\bibitem [{\citenamefont {Pham-Van-Diep}\ \emph {et~al.}(1991)\citenamefont
  {Pham-Van-Diep}, \citenamefont {Erwin},\ and\ \citenamefont
  {Muntz}}]{PEM1991}%
  \BibitemOpen
  \bibfield  {author} {\bibinfo {author} {\bibfnamefont {G.~C.}\ \bibnamefont
  {Pham-Van-Diep}}, \bibinfo {author} {\bibfnamefont {D.~A.}\ \bibnamefont
  {Erwin}},\ and\ \bibinfo {author} {\bibfnamefont {E.~P.}\ \bibnamefont
  {Muntz}},\ }\bibfield  {title} {\bibinfo {title} {Testing continuum
  descriptions of low-{M}ach-number shock structures},\ }\href
  {https://doi.org/10.1017/S0022112091003749} {\bibfield  {journal} {\bibinfo
  {journal} {J.~Fluid Mech.}\ }\textbf {\bibinfo {volume} {232}},\ \bibinfo
  {pages} {403} (\bibinfo {year} {1991})}\BibitemShut {NoStop}%
\bibitem [{\citenamefont {Alsmeyer}(1976)}]{Alsmeyer1976}%
  \BibitemOpen
  \bibfield  {author} {\bibinfo {author} {\bibfnamefont {H.}~\bibnamefont
  {Alsmeyer}},\ }\bibfield  {title} {\bibinfo {title} {Density profiles in
  argon and nitrogen shock waves measured by the absorption of an electron
  beam},\ }\href@noop {} {\bibfield  {journal} {\bibinfo  {journal} {J.~Fluid
  Mech.}\ }\textbf {\bibinfo {volume} {74}},\ \bibinfo {pages} {497} (\bibinfo
  {year} {1976})}\BibitemShut {NoStop}%
\bibitem [{\citenamefont {Linzer}\ and\ \citenamefont {Hornig}(1963)}]{LH1963}%
  \BibitemOpen
  \bibfield  {author} {\bibinfo {author} {\bibfnamefont {M.}~\bibnamefont
  {Linzer}}\ and\ \bibinfo {author} {\bibfnamefont {D.~F.}\ \bibnamefont
  {Hornig}},\ }\bibfield  {title} {\bibinfo {title} {Structure of shock fronts
  in argon and nitrogen},\ }\href {https://doi.org/10.1063/1.1711007}
  {\bibfield  {journal} {\bibinfo  {journal} {Phys. Fluids}\ }\textbf {\bibinfo
  {volume} {6}},\ \bibinfo {pages} {1661} (\bibinfo {year} {1963})}\BibitemShut
  {NoStop}%
\bibitem [{\citenamefont {{Camac}}(1965)}]{Camac1965}%
  \BibitemOpen
  \bibfield  {author} {\bibinfo {author} {\bibfnamefont {M.}~\bibnamefont
  {{Camac}}},\ }\href@noop {} {\emph {\bibinfo {title} {In Proceedings of the
  Fourth International Symposium on Rarefied Gas Dynamics}}},\ edited by\
  \bibinfo {editor} {\bibfnamefont {J.~H.}\ \bibnamefont {de~Leeuw}},\
  Vol.~\bibinfo {volume} {1}\ (\bibinfo  {publisher} {Academic Press},\
  \bibinfo {year} {1965})\ p.\ \bibinfo {pages} {240}\BibitemShut {NoStop}%
\bibitem [{\citenamefont {{Schultz-Grunow}}\ and\ \citenamefont
  {{Frohn}}(1965)}]{Sf1965}%
  \BibitemOpen
  \bibfield  {author} {\bibinfo {author} {\bibfnamefont {F.}~\bibnamefont
  {{Schultz-Grunow}}}\ and\ \bibinfo {author} {\bibfnamefont {A.}~\bibnamefont
  {{Frohn}}},\ }\href@noop {} {\emph {\bibinfo {title} {In Proceedings of the
  Fourth International Symposium on Rarefied Gas Dynamics}}},\ edited by\
  \bibinfo {editor} {\bibfnamefont {J.~H.}\ \bibnamefont {de~Leeuw}},\
  Vol.~\bibinfo {volume} {1}\ (\bibinfo  {publisher} {Academic Press},\
  \bibinfo {year} {1965})\ p.\ \bibinfo {pages} {250}\BibitemShut {NoStop}%
\bibitem [{\citenamefont {Russell}(1965)}]{Russell1965}%
  \BibitemOpen
  \bibfield  {author} {\bibinfo {author} {\bibfnamefont {D.}~\bibnamefont
  {Russell}},\ }\href@noop {} {\emph {\bibinfo {title} {In Proceedings of the
  Fourth International Symposium on Rarefied Gas Dynamics}}},\ edited by\
  \bibinfo {editor} {\bibfnamefont {J.~H.}\ \bibnamefont {de~Leeuw}},\
  Vol.~\bibinfo {volume} {1}\ (\bibinfo  {publisher} {Academic Press},\
  \bibinfo {year} {1965})\ p.\ \bibinfo {pages} {265}\BibitemShut {NoStop}%
\bibitem [{\citenamefont {Robben}\ and\ \citenamefont {Talbot}(1966)}]{RT1966}%
  \BibitemOpen
  \bibfield  {author} {\bibinfo {author} {\bibfnamefont {F.}~\bibnamefont
  {Robben}}\ and\ \bibinfo {author} {\bibfnamefont {L.}~\bibnamefont
  {Talbot}},\ }\bibfield  {title} {\bibinfo {title} {Measurement of shock wave
  thickness by the electron beam fluorescence method},\ }\href
  {https://doi.org/10.1063/1.1761728} {\bibfield  {journal} {\bibinfo
  {journal} {Phys. Fluids}\ }\textbf {\bibinfo {volume} {9}},\ \bibinfo {pages}
  {633} (\bibinfo {year} {1966})}\BibitemShut {NoStop}%
\bibitem [{\citenamefont {Schmidt}(1969)}]{Schmidt1969}%
  \BibitemOpen
  \bibfield  {author} {\bibinfo {author} {\bibfnamefont {B.}~\bibnamefont
  {Schmidt}},\ }\bibfield  {title} {\bibinfo {title} {Electron beam density
  measurements in shock waves in argon},\ }\href
  {https://doi.org/10.1017/S0022112069002229} {\bibfield  {journal} {\bibinfo
  {journal} {J.~Fluid Mech.}\ }\textbf {\bibinfo {volume} {39}},\ \bibinfo
  {pages} {361} (\bibinfo {year} {1969})}\BibitemShut {NoStop}%
\bibitem [{\citenamefont {Rieutord}(1970)}]{Rieutord1970}%
  \BibitemOpen
  \bibfield  {author} {\bibinfo {author} {\bibfnamefont {E.}~\bibnamefont
  {Rieutord}},\ }\emph {\bibinfo {title} {Contribution à l’étude de
  l’onde de choc primaire en tube à choc}},\ \href@noop {} {Ph.D. thesis},\
  \bibinfo  {school} {Institut National des Sciences Appliquées de Lyon}
  (\bibinfo {year} {1970})\BibitemShut {NoStop}%
\bibitem [{\citenamefont {Garen}\ \emph {et~al.}(1974)\citenamefont {Garen},
  \citenamefont {Synofzik},\ and\ \citenamefont {Frohn}}]{Garenetal1974}%
  \BibitemOpen
  \bibfield  {author} {\bibinfo {author} {\bibfnamefont {W.}~\bibnamefont
  {Garen}}, \bibinfo {author} {\bibfnamefont {R.}~\bibnamefont {Synofzik}},\
  and\ \bibinfo {author} {\bibfnamefont {A.}~\bibnamefont {Frohn}},\ }\bibfield
   {title} {\bibinfo {title} {Shock tube for generating weak shock waves},\
  }\href {https://doi.org/10.2514/3.49425} {\bibfield  {journal} {\bibinfo
  {journal} {AIAA Journal}\ }\textbf {\bibinfo {volume} {12}},\ \bibinfo
  {pages} {1132} (\bibinfo {year} {1974})}\BibitemShut {NoStop}%
\bibitem [{\citenamefont {Sone}(2000)}]{Sone2000}%
  \BibitemOpen
  \bibfield  {author} {\bibinfo {author} {\bibfnamefont {Y.}~\bibnamefont
  {Sone}},\ }\bibfield  {title} {\bibinfo {title} {Flows induced by temperature
  fields in a rarefied gas and their ghost effect on the behavior of a gas in
  the continuum limit},\ }\href
  {https://doi.org/10.1146/annurev.fluid.32.1.779} {\bibfield  {journal}
  {\bibinfo  {journal} {Annu. Rev. Fluid Mech.}\ }\textbf {\bibinfo {volume}
  {32}},\ \bibinfo {pages} {779} (\bibinfo {year} {2000})}\BibitemShut
  {NoStop}%
\bibitem [{\citenamefont {Durst}(2008)}]{Durst2008}%
  \BibitemOpen
  \bibfield  {author} {\bibinfo {author} {\bibfnamefont {F.}~\bibnamefont
  {Durst}},\ }\href@noop {} {\emph {\bibinfo {title} {Fluid Mechanics: An
  Introduction to the Theory of Fluid Flows}}}\ (\bibinfo  {publisher}
  {Springer, Berlin},\ \bibinfo {year} {2008})\BibitemShut {NoStop}%
\bibitem [{\citenamefont {Stokes}(1845)}]{Stokes1845}%
  \BibitemOpen
  \bibfield  {author} {\bibinfo {author} {\bibfnamefont {G.~G.}\ \bibnamefont
  {Stokes}},\ }\bibfield  {title} {\bibinfo {title} {On the theories of the
  internal friction of fluids in motion, and of the equilibrium and motion of
  elastic solids},\ }\href@noop {} {\bibfield  {journal} {\bibinfo  {journal}
  {Trans. Camb. Phil. Soc.}\ }\textbf {\bibinfo {volume} {8}},\ \bibinfo
  {pages} {287} (\bibinfo {year} {1845})}\BibitemShut {NoStop}%
\bibitem [{\citenamefont {Boltzmann}(1878)}]{Boltzmann1878}%
  \BibitemOpen
  \bibfield  {author} {\bibinfo {author} {\bibfnamefont {L.}~\bibnamefont
  {Boltzmann}},\ }\bibfield  {title} {\bibinfo {title} {Zur theorie der
  elastischen nachwirkung},\ }\href@noop {} {\bibfield  {journal} {\bibinfo
  {journal} {Annalen der Physik}\ }\textbf {\bibinfo {volume} {241}},\ \bibinfo
  {pages} {430} (\bibinfo {year} {1878})}\BibitemShut {NoStop}%
\bibitem [{\citenamefont {Bird}(1970{\natexlab{a}})}]{Bird1970a}%
  \BibitemOpen
  \bibfield  {author} {\bibinfo {author} {\bibfnamefont {G.~A.}\ \bibnamefont
  {Bird}},\ }\bibfield  {title} {\bibinfo {title} {Aspects of the structure of
  strong shock waves},\ }\href {https://doi.org/10.1063/1.1693047} {\bibfield
  {journal} {\bibinfo  {journal} {Phys. Fluids}\ }\textbf {\bibinfo {volume}
  {13}},\ \bibinfo {pages} {1172} (\bibinfo {year}
  {1970}{\natexlab{a}})}\BibitemShut {NoStop}%
\bibitem [{\citenamefont {Bird}(1970{\natexlab{b}})}]{Bird1970b}%
  \BibitemOpen
  \bibfield  {author} {\bibinfo {author} {\bibfnamefont {G.~A.}\ \bibnamefont
  {Bird}},\ }\bibfield  {title} {\bibinfo {title} {Direct simulation and the
  {B}oltzmann equation},\ }\href {https://doi.org/10.1063/1.1692849} {\bibfield
   {journal} {\bibinfo  {journal} {Phys. Fluids}\ }\textbf {\bibinfo {volume}
  {13}},\ \bibinfo {pages} {2676} (\bibinfo {year}
  {1970}{\natexlab{b}})}\BibitemShut {NoStop}%
\bibitem [{\citenamefont {Grad}(1949)}]{Grad1949}%
  \BibitemOpen
  \bibfield  {author} {\bibinfo {author} {\bibfnamefont {H.}~\bibnamefont
  {Grad}},\ }\bibfield  {title} {\bibinfo {title} {On the kinetic theory of
  rarefied gases},\ }\href@noop {} {\bibfield  {journal} {\bibinfo  {journal}
  {Commun.~Pure Appl.~Math.}\ }\textbf {\bibinfo {volume} {2}},\ \bibinfo
  {pages} {331} (\bibinfo {year} {1949})}\BibitemShut {NoStop}%
\bibitem [{\citenamefont {Chapman}\ and\ \citenamefont
  {Cowling}(1970)}]{CC1970}%
  \BibitemOpen
  \bibfield  {author} {\bibinfo {author} {\bibfnamefont {S.}~\bibnamefont
  {Chapman}}\ and\ \bibinfo {author} {\bibfnamefont {T.~G.}\ \bibnamefont
  {Cowling}},\ }\href@noop {} {\emph {\bibinfo {title} {The {M}athematical
  {T}heory of {N}on-uniform {G}ases}}}\ (\bibinfo  {publisher} {Cambridge
  university press},\ \bibinfo {year} {1970})\BibitemShut {NoStop}%
\bibitem [{\citenamefont {Cercignani}(1975)}]{Cercignani1975}%
  \BibitemOpen
  \bibfield  {author} {\bibinfo {author} {\bibfnamefont {C.}~\bibnamefont
  {Cercignani}},\ }\href@noop {} {\emph {\bibinfo {title} {Theory and
  application of the Boltzmann equation}}}\ (\bibinfo  {publisher} {Scottish
  Academic Press},\ \bibinfo {year} {1975})\BibitemShut {NoStop}%
\bibitem [{\citenamefont {Gorban}\ and\ \citenamefont {Karlin}(1994)}]{GK1994}%
  \BibitemOpen
  \bibfield  {author} {\bibinfo {author} {\bibfnamefont {A.~N.}\ \bibnamefont
  {Gorban}}\ and\ \bibinfo {author} {\bibfnamefont {I.~V.}\ \bibnamefont
  {Karlin}},\ }\bibfield  {title} {\bibinfo {title} {General approach to
  constructing models of the boltzmann equation},\ }\href
  {https://doi.org/https://doi.org/10.1016/0378-4371(94)90314-X} {\bibfield
  {journal} {\bibinfo  {journal} {Physica A}\ }\textbf {\bibinfo {volume}
  {206}},\ \bibinfo {pages} {401} (\bibinfo {year} {1994})}\BibitemShut
  {NoStop}%
\bibitem [{\citenamefont {Struchtrup}(2005)}]{HS2005}%
  \BibitemOpen
  \bibfield  {author} {\bibinfo {author} {\bibfnamefont {H.}~\bibnamefont
  {Struchtrup}},\ }\href@noop {} {\emph {\bibinfo {title} {Macroscopic
  Transport Equations for Rarefied Gas Flows}}}\ (\bibinfo  {publisher}
  {Springer},\ \bibinfo {year} {2005})\BibitemShut {NoStop}%
\bibitem [{\citenamefont {Enskog}(1917)}]{Enskog1917}%
  \BibitemOpen
  \bibfield  {author} {\bibinfo {author} {\bibfnamefont {D.}~\bibnamefont
  {Enskog}},\ }\emph {\bibinfo {title} {Theorie der {V}org{\"a}nge in {M}assing
  {V}erdumten {G}asen}},\ \href@noop {} {Ph.D. thesis},\ \bibinfo  {school}
  {Ph. D thesis, University of Uppsala, Sweden} (\bibinfo {year}
  {1917})\BibitemShut {NoStop}%
\bibitem [{\citenamefont {Reddy}\ and\ \citenamefont {Ganesan}(2019)}]{RG2019}%
  \BibitemOpen
  \bibfield  {author} {\bibinfo {author} {\bibfnamefont {M.~H.~L.}\
  \bibnamefont {Reddy}}\ and\ \bibinfo {author} {\bibfnamefont
  {S.}~\bibnamefont {Ganesan}},\ }\bibfield  {title} {\bibinfo {title}
  {Thirty-five moment theory for dilute smooth hard sphere gases},\ }\href
  {https://doi.org/10.1063/1.5119615} {\bibfield  {journal} {\bibinfo
  {journal} {AIP Conf. Proc.}\ }\textbf {\bibinfo {volume} {2132}},\ \bibinfo
  {pages} {120002} (\bibinfo {year} {2019})}\BibitemShut {NoStop}%
\bibitem [{\citenamefont {Kogan}(1969)}]{Kogan1969}%
  \BibitemOpen
  \bibfield  {author} {\bibinfo {author} {\bibfnamefont {M.~N.}\ \bibnamefont
  {Kogan}},\ }\href@noop {} {\emph {\bibinfo {title} {Rarefied Gas Dynamics}}}\
  (\bibinfo  {publisher} {Plenum Press, New York},\ \bibinfo {year}
  {1969})\BibitemShut {NoStop}%
\bibitem [{\citenamefont {Levermore}(1996)}]{LM1996}%
  \BibitemOpen
  \bibfield  {author} {\bibinfo {author} {\bibfnamefont {C.~D.}\ \bibnamefont
  {Levermore}},\ }\bibfield  {title} {\bibinfo {title} {Moment closure
  hierarchies for kinetic theories},\ }\href@noop {} {\bibfield  {journal}
  {\bibinfo  {journal} {Journal of Statistical Physics}\ }\textbf {\bibinfo
  {volume} {83}},\ \bibinfo {pages} {1021} (\bibinfo {year}
  {1996})}\BibitemShut {NoStop}%
\bibitem [{\citenamefont {Reddy}\ \emph {et~al.}(2014)\citenamefont {Reddy},
  \citenamefont {Ansumali},\ and\ \citenamefont {Alam}}]{Reddyetal2014}%
  \BibitemOpen
  \bibfield  {author} {\bibinfo {author} {\bibfnamefont {M.~H.~L.}\
  \bibnamefont {Reddy}}, \bibinfo {author} {\bibfnamefont {S.}~\bibnamefont
  {Ansumali}},\ and\ \bibinfo {author} {\bibfnamefont {M.}~\bibnamefont
  {Alam}},\ }\bibfield  {title} {\bibinfo {title} {Shock waves in a dilute
  granular gas},\ }\href {https://doi.org/10.1063/1.4902632} {\bibfield
  {journal} {\bibinfo  {journal} {AIP Conf. Proc.}\ }\textbf {\bibinfo {volume}
  {1628}},\ \bibinfo {pages} {480} (\bibinfo {year} {2014})}\BibitemShut
  {NoStop}%
\bibitem [{\citenamefont {Weiss}(1995)}]{Weiss1995}%
  \BibitemOpen
  \bibfield  {author} {\bibinfo {author} {\bibfnamefont {W.}~\bibnamefont
  {Weiss}},\ }\bibfield  {title} {\bibinfo {title} {Continuous shock structure
  in extended thermodynamics},\ }\href@noop {} {\bibfield  {journal} {\bibinfo
  {journal} {Phys. Rev. E}\ }\textbf {\bibinfo {volume} {52}},\ \bibinfo
  {pages} {R5760} (\bibinfo {year} {1995})}\BibitemShut {NoStop}%
\bibitem [{\citenamefont {Torrilhon}\ and\ \citenamefont
  {Struchtrup}(2004)}]{TS2004}%
  \BibitemOpen
  \bibfield  {author} {\bibinfo {author} {\bibfnamefont {M.}~\bibnamefont
  {Torrilhon}}\ and\ \bibinfo {author} {\bibfnamefont {H.}~\bibnamefont
  {Struchtrup}},\ }\bibfield  {title} {\bibinfo {title} {Regularized 13-moment
  equations: shock structure calculations and comparison to burnett models},\
  }\href {https://doi.org/10.1017/S0022112004009917} {\bibfield  {journal}
  {\bibinfo  {journal} {J.~Fluid Mech.}\ }\textbf {\bibinfo {volume} {513}},\
  \bibinfo {pages} {171} (\bibinfo {year} {2004})}\BibitemShut {NoStop}%
\bibitem [{\citenamefont {Lumpkin}\ and\ \citenamefont
  {Chapman}(1992)}]{LC1992}%
  \BibitemOpen
  \bibfield  {author} {\bibinfo {author} {\bibfnamefont {F.~E.}\ \bibnamefont
  {Lumpkin}}\ and\ \bibinfo {author} {\bibfnamefont {D.~R.}\ \bibnamefont
  {Chapman}},\ }\bibfield  {title} {\bibinfo {title} {Accuracy of the burnett
  equations for hypersonic real gas flows},\ }\href
  {https://doi.org/10.2514/3.377} {\bibfield  {journal} {\bibinfo  {journal}
  {J. Thermophys. Heat Transfer}\ }\textbf {\bibinfo {volume} {6}},\ \bibinfo
  {pages} {419} (\bibinfo {year} {1992})}\BibitemShut {NoStop}%
\bibitem [{\citenamefont {Balakrishnan}(2004)}]{Balakrishnan2004}%
  \BibitemOpen
  \bibfield  {author} {\bibinfo {author} {\bibfnamefont {R.}~\bibnamefont
  {Balakrishnan}},\ }\bibfield  {title} {\bibinfo {title} {An approach to
  entropy consistency in second-order hydrodynamic equations},\ }\href@noop {}
  {\bibfield  {journal} {\bibinfo  {journal} {J.~Fluid Mech.}\ }\textbf
  {\bibinfo {volume} {503}},\ \bibinfo {pages} {201} (\bibinfo {year}
  {2004})}\BibitemShut {NoStop}%
\bibitem [{\citenamefont {Jin}\ and\ \citenamefont {Slemrod}(2001)}]{JS2001}%
  \BibitemOpen
  \bibfield  {author} {\bibinfo {author} {\bibfnamefont {S.}~\bibnamefont
  {Jin}}\ and\ \bibinfo {author} {\bibfnamefont {M.}~\bibnamefont {Slemrod}},\
  }\bibfield  {title} {\bibinfo {title} {Regularization of the {B}urnett
  equations via relaxation},\ }\href@noop {} {\bibfield  {journal} {\bibinfo
  {journal} {J.~Stat. Phys.}\ }\textbf {\bibinfo {volume} {103}},\ \bibinfo
  {pages} {1009} (\bibinfo {year} {2001})}\BibitemShut {NoStop}%
\bibitem [{\citenamefont {Reddy}\ and\ \citenamefont {Alam}(2020)}]{RA2020}%
  \BibitemOpen
  \bibfield  {author} {\bibinfo {author} {\bibfnamefont {M.~H.~L.}\
  \bibnamefont {Reddy}}\ and\ \bibinfo {author} {\bibfnamefont
  {M.}~\bibnamefont {Alam}},\ }\bibfield  {title} {\bibinfo {title}
  {Regularized extended-hydrodynamic equations for a rarefied granular gas and
  the plane shock waves},\ }\href
  {https://doi.org/10.1103/PhysRevFluids.5.044302} {\bibfield  {journal}
  {\bibinfo  {journal} {Phys. Rev. Fluids}\ }\textbf {\bibinfo {volume} {5}},\
  \bibinfo {pages} {044302} (\bibinfo {year} {2020})}\BibitemShut {NoStop}%
\bibitem [{\citenamefont {Reddy}\ and\ \citenamefont
  {Alam}(2016{\natexlab{b}})}]{RA2016a}%
  \BibitemOpen
  \bibfield  {author} {\bibinfo {author} {\bibfnamefont {M.~H.~L.}\
  \bibnamefont {Reddy}}\ and\ \bibinfo {author} {\bibfnamefont
  {M.}~\bibnamefont {Alam}},\ }\href@noop {} {\bibinfo {title} {Regularized
  moment equations and plane shock waves for a rarefied granular gas}},\
  \bibinfo {howpublished} {Bull.~American Phys.~Soc.} (\bibinfo {year}
  {2016}{\natexlab{b}})\BibitemShut {NoStop}%
\bibitem [{\citenamefont {Woods}(1993)}]{Woods1993}%
  \BibitemOpen
  \bibfield  {author} {\bibinfo {author} {\bibfnamefont {L.~C.}\ \bibnamefont
  {Woods}},\ }\href@noop {} {\emph {\bibinfo {title} {An Introduction to the
  Kinetic Theory of Gases and Magnetoplasmas.}}}\ (\bibinfo  {publisher}
  {Oxford University Press, Oxford},\ \bibinfo {year} {1993})\BibitemShut
  {NoStop}%
\bibitem [{\citenamefont {Paolucci}\ and\ \citenamefont
  {Paolucci}(2018)}]{Paolucci2018}%
  \BibitemOpen
  \bibfield  {author} {\bibinfo {author} {\bibfnamefont {S.}~\bibnamefont
  {Paolucci}}\ and\ \bibinfo {author} {\bibfnamefont {C.}~\bibnamefont
  {Paolucci}},\ }\bibfield  {title} {\bibinfo {title} {A second-order continuum
  theory of fluids},\ }\href {https://doi.org/10.1017/jfm.2018.291} {\bibfield
  {journal} {\bibinfo  {journal} {J.~Fluid Mech.}\ }\textbf {\bibinfo {volume}
  {846}},\ \bibinfo {pages} {686} (\bibinfo {year} {2018})}\BibitemShut
  {NoStop}%
\bibitem [{\citenamefont {Velasco}\ and\ \citenamefont
  {Uribe}(2019)}]{VelascoUribe2019}%
  \BibitemOpen
  \bibfield  {author} {\bibinfo {author} {\bibfnamefont {R.~M.}\ \bibnamefont
  {Velasco}}\ and\ \bibinfo {author} {\bibfnamefont {F.~J.}\ \bibnamefont
  {Uribe}},\ }\bibfield  {title} {\bibinfo {title} {Shock-wave structure
  according to a linear irreversible thermodynamic model},\ }\href
  {https://doi.org/10.1103/PhysRevE.99.023114} {\bibfield  {journal} {\bibinfo
  {journal} {Phys. Rev. E}\ }\textbf {\bibinfo {volume} {99}},\ \bibinfo
  {pages} {023114} (\bibinfo {year} {2019})}\BibitemShut {NoStop}%
\bibitem [{\citenamefont {Reddy}\ and\ \citenamefont {Dadzie}(2020)}]{RD2020}%
  \BibitemOpen
  \bibfield  {author} {\bibinfo {author} {\bibfnamefont {M.~H.~L.}\
  \bibnamefont {Reddy}}\ and\ \bibinfo {author} {\bibfnamefont {S.~K.}\
  \bibnamefont {Dadzie}},\ }\bibfield  {title} {\bibinfo {title}
  {Reinterpreting shock wave structure predictions using the {N}avier-{S}tokes
  equations},\ }\href {https://doi.org/10.1007/s00193-020-00952-1} {\bibfield
  {journal} {\bibinfo  {journal} {Shock Waves}\ }\textbf {\bibinfo {volume}
  {30}},\ \bibinfo {pages} {513} (\bibinfo {year} {2020})}\BibitemShut
  {NoStop}%
\bibitem [{\citenamefont {Jadhav}\ and\ \citenamefont
  {Agrawal}(2020)}]{JadhavAgrawal2020}%
  \BibitemOpen
  \bibfield  {author} {\bibinfo {author} {\bibfnamefont {R.~S.}\ \bibnamefont
  {Jadhav}}\ and\ \bibinfo {author} {\bibfnamefont {A.}~\bibnamefont
  {Agrawal}},\ }\bibfield  {title} {\bibinfo {title} {Strong shock as a
  stringent test for onsager-burnett equations},\ }\href
  {https://doi.org/10.1103/PhysRevE.102.063111} {\bibfield  {journal} {\bibinfo
   {journal} {Phys. Rev. E}\ }\textbf {\bibinfo {volume} {102}},\ \bibinfo
  {pages} {063111} (\bibinfo {year} {2020})}\BibitemShut {NoStop}%
\bibitem [{\citenamefont {Dadzie}(2013)}]{Dadzie2013}%
  \BibitemOpen
  \bibfield  {author} {\bibinfo {author} {\bibfnamefont {S.~K.}\ \bibnamefont
  {Dadzie}},\ }\bibfield  {title} {\bibinfo {title} {A thermo-mechanically
  consistent {B}urnett regime continuum flow equation without
  {C}hapman-{E}nskog expansion},\ }\href@noop {} {\bibfield  {journal}
  {\bibinfo  {journal} {J.~Fluid Mech.}\ }\textbf {\bibinfo {volume} {716}},\
  \bibinfo {pages} {R6} (\bibinfo {year} {2013})}\BibitemShut {NoStop}%
\bibitem [{\citenamefont {Brenner}(2012)}]{Brenner2012}%
  \BibitemOpen
  \bibfield  {author} {\bibinfo {author} {\bibfnamefont {H.}~\bibnamefont
  {Brenner}},\ }\bibfield  {title} {\bibinfo {title} {Beyond
  {N}avier-{S}tokes},\ }\href@noop {} {\bibfield  {journal} {\bibinfo
  {journal} {Int. J. Eng. Sci.}\ }\textbf {\bibinfo {volume} {54}},\ \bibinfo
  {pages} {67} (\bibinfo {year} {2012})}\BibitemShut {NoStop}%
\bibitem [{\citenamefont {Dadzie}\ and\ \citenamefont
  {Christou}(2016)}]{DadzieChariton2016}%
  \BibitemOpen
  \bibfield  {author} {\bibinfo {author} {\bibfnamefont {S.~K.}\ \bibnamefont
  {Dadzie}}\ and\ \bibinfo {author} {\bibfnamefont {C.}~\bibnamefont
  {Christou}},\ }\bibfield  {title} {\bibinfo {title} {Bi-velocity gas dynamics
  of a micro lid-driven cavity heat transfer subject to forced convection},\
  }\href
  {https://doi.org/https://doi.org/10.1016/j.icheatmasstransfer.2016.09.006}
  {\bibfield  {journal} {\bibinfo  {journal} {International Communications in
  Heat and Mass Transfer}\ }\textbf {\bibinfo {volume} {78}},\ \bibinfo {pages}
  {175} (\bibinfo {year} {2016})}\BibitemShut {NoStop}%
\bibitem [{\citenamefont {Christou}\ and\ \citenamefont
  {Dadzie}(2017)}]{CharitonDadzie2017}%
  \BibitemOpen
  \bibfield  {author} {\bibinfo {author} {\bibfnamefont {C.}~\bibnamefont
  {Christou}}\ and\ \bibinfo {author} {\bibfnamefont {S.~K.}\ \bibnamefont
  {Dadzie}},\ }\bibfield  {title} {\bibinfo {title} {{An investigation of heat
  transfer in a cavity flow in the non-continuum regime}},\ }\bibfield
  {journal} {\bibinfo  {journal} {J. Heat Transfer.}\ }\textbf {\bibinfo
  {volume} {139}},\ \href {https://doi.org/10.1115/1.4036340}
  {10.1115/1.4036340} (\bibinfo {year} {2017})\BibitemShut {NoStop}%
\bibitem [{\citenamefont {Dadzie}\ and\ \citenamefont
  {Reese}(2010)}]{DadzieReese2010}%
  \BibitemOpen
  \bibfield  {author} {\bibinfo {author} {\bibfnamefont {S.~K.}\ \bibnamefont
  {Dadzie}}\ and\ \bibinfo {author} {\bibfnamefont {J.~M.}\ \bibnamefont
  {Reese}},\ }\bibfield  {title} {\bibinfo {title} {A volume-based hydrodynamic
  approach to sound wave propagation in a monatomic gas},\ }\href@noop {}
  {\bibfield  {journal} {\bibinfo  {journal} {Phys. Fluids}\ }\textbf {\bibinfo
  {volume} {22}},\ \bibinfo {pages} {016103} (\bibinfo {year}
  {2010})}\BibitemShut {NoStop}%
\bibitem [{\citenamefont {Reddy}\ \emph {et~al.}(2019)\citenamefont {Reddy},
  \citenamefont {Dadzie}, \citenamefont {Ocone}, \citenamefont {Borg},\ and\
  \citenamefont {Reese}}]{Reddyetal2019}%
  \BibitemOpen
  \bibfield  {author} {\bibinfo {author} {\bibfnamefont {M.~H.~L.}\
  \bibnamefont {Reddy}}, \bibinfo {author} {\bibfnamefont {S.~K.}\ \bibnamefont
  {Dadzie}}, \bibinfo {author} {\bibfnamefont {R.}~\bibnamefont {Ocone}},
  \bibinfo {author} {\bibfnamefont {M.~K.}\ \bibnamefont {Borg}},\ and\
  \bibinfo {author} {\bibfnamefont {J.~M.}\ \bibnamefont {Reese}},\ }\bibfield
  {title} {\bibinfo {title} {Recasting {N}avier{\textendash}{S}tokes
  equations},\ }\href {https://doi.org/10.1088/2399-6528/ab4b86} {\bibfield
  {journal} {\bibinfo  {journal} {J. Phys. Commun.}\ }\textbf {\bibinfo
  {volume} {3}},\ \bibinfo {pages} {105009} (\bibinfo {year}
  {2019})}\BibitemShut {NoStop}%
\bibitem [{\citenamefont {Mackenzie}(2006)}]{Mackenzie2006}%
  \BibitemOpen
  \bibfield  {author} {\bibinfo {author} {\bibfnamefont {N.~S.}\ \bibnamefont
  {Mackenzie}},\ }\emph {\bibinfo {title} {Stability of the {B}eagle 2 {M}ars
  {E}xplorer}},\ \href@noop {} {Ph.D. thesis},\ \bibinfo  {school} {University
  of Melbourne} (\bibinfo {year} {2006})\BibitemShut {NoStop}%
\bibitem [{\citenamefont {Dadzie}\ and\ \citenamefont {Reddy}(2020)}]{DR2020}%
  \BibitemOpen
  \bibfield  {author} {\bibinfo {author} {\bibfnamefont {S.~K.}\ \bibnamefont
  {Dadzie}}\ and\ \bibinfo {author} {\bibfnamefont {M.~H.~L.}\ \bibnamefont
  {Reddy}},\ }\bibfield  {title} {\bibinfo {title} {Recasting {N}avier-{S}tokes
  equations: shock wave structure description},\ }\href
  {https://doi.org/10.1063/5.0026758} {\bibfield  {journal} {\bibinfo
  {journal} {AIP Conf. Proc.}\ }\textbf {\bibinfo {volume} {2293}},\ \bibinfo
  {pages} {050005} (\bibinfo {year} {2020})}\BibitemShut {NoStop}%
\bibitem [{\citenamefont {Reddy}\ and\ \citenamefont {Dadzie}(2021)}]{RD2021}%
  \BibitemOpen
  \bibfield  {author} {\bibinfo {author} {\bibfnamefont {M.~H.~L.}\
  \bibnamefont {Reddy}}\ and\ \bibinfo {author} {\bibfnamefont {S.~K.}\
  \bibnamefont {Dadzie}},\ }\bibfield  {title} {\bibinfo {title} {Accurate
  constitutive relations for shock wave structures in gases},\ }\href
  {https://doi.org/10.2495/MPF210141} {\bibfield  {journal} {\bibinfo
  {journal} {WIT Trans. Eng. Sci.}\ }\textbf {\bibinfo {volume} {132}},\
  \bibinfo {pages} {165} (\bibinfo {year} {2021})}\BibitemShut {NoStop}%
\bibitem [{\citenamefont {Schrock}(2005)}]{Schrock2005}%
  \BibitemOpen
  \bibfield  {author} {\bibinfo {author} {\bibfnamefont {C.~R.}\ \bibnamefont
  {Schrock}},\ }\href@noop {} {\bibinfo {title} {Entropy generation as a means
  of examining continuum breakdown}},\ \bibinfo {howpublished} {Master’s
  thesis, Air Force Institute of Technology, Wright-Patterson AFB, Ohio}
  (\bibinfo {year} {2005})\BibitemShut {NoStop}%
\end{thebibliography}

\providecommand{\noopsort}[1]{}\providecommand{\singleletter}[1]{#1}%
%

\end{document}